\documentclass[draftclsnofoot,onecolumn,11pt]{IEEEtran}
\usepackage{amsmath,amsfonts,amssymb}
\usepackage{graphicx}
\usepackage{cite}

\newcommand{\mQ}{\ensuremath{\mathbf Q}}

\newcommand{\vy}{\ensuremath{\mathbf y}}
\newcommand{\vx}{\ensuremath{\mathbf x}}
\newcommand{\vs}{\ensuremath{\mathbf s}}
\newcommand{\vv}{\ensuremath{\mathbf v}}

\newcommand{\vq}{\ensuremath{\mathbf q}}

\newcommand{\vw}{\ensuremath{\mathbf w}}
\newcommand{\ball}[2]{\ensuremath{\mathcal{B_{#1}({#2})}}}

\newcommand{\sS}{\ensuremath{\mathcal{S}}}
\newcommand{\Reals}{\ensuremath{\mathbb{R}}}

\newcommand{\sinc}{\ensuremath{\mathrm{sinc}}}

\newtheorem{theorem}{Theorem}[section]
\newtheorem{lemma}[theorem]{Lemma}

\newtheorem{corollary}[theorem]{Corollary}

\newcommand{\qed}{\nobreak \ifvmode \relax \else
      \ifdim\lastskip<1.5em \hskip-\lastskip
      \hskip1.5em plus0em minus0.5em \fi \nobreak
      \vrule height0.75em width0.5em depth0.25em\fi}

\begin{document}
\title{Universal Rate-Efficient Scalar Quantization} 
\author{Petros T. Boufounos,~\IEEEmembership{Member,~IEEE}}

\maketitle 
\begin{abstract}
Scalar quantization is the most practical and straightforward approach
to signal quantization. However, it has been shown that scalar
quantization of oversampled or Compressively Sensed signals can be
inefficient in terms of the rate-distortion trade-off, especially as
the oversampling rate or the sparsity of the signal increases. In this
paper, we modify the scalar quantizer to have discontinuous
quantization regions. We demonstrate that with this modification it is
possible to achieve exponential decay of the quantization error as a
function of the oversampling rate instead of the quadratic decay
exhibited by current approaches. Our approach is universal in the
sense that prior knowledge of the signal model is not necessary in the
quantizer design, only in the reconstruction. Thus, we demonstrate
that it is possible to reduce the quantization error by incorporating
side information on the acquired signal, such as sparse signal models
or signal similarity with known signals. In doing so, we establish a
relationship between quantization performance and the Kolmogorov
entropy of the signal model.
\end{abstract}

\begin{IEEEkeywords}
scalar quantization, randomization, randomized embedding,
oversampling, robustness
\end{IEEEkeywords}

\IEEEpeerreviewmaketitle

\section{Introduction}
\label{sec:intro}
\IEEEPARstart{I}{n} order to digitize a signal, two discretization
steps are necessary: sampling (or measurement) and quantization. The
first step, sampling, computes linear functions of the signal, such as
the signal's instantaneous value or the signal's inner product with a
measurement vector. The second step, quantization, maps the
continuous-valued measurements of the signal to a set of discrete
values, usually referred to as quantization points. Overall, these two
discretization steps do not preserve all the information in the analog
signal.

The sampling step of the discretization can be designed to preserve
all the information in the signal. Several sampling results
demonstrate that as long as sufficiently many samples are obtained
given the class of the signal sampled it is possible to exactly
recover a signal from its samples. The most celebrated sampling result
is the Nyquist sampling theorem which dictates that uniform sampling
at a frequency at least twice the bandwidth of a signal is sufficient
to recover the signal using simple bandlimited interpolation. More
recently, Compressive Sensing theory has demonstrated that it is also
possible to recover a sparse signal from samples approximately at its
sparsity rate, rather than its Nyquist rate or the rate implied by the
dimension of the signal.

Unfortunately, the quantization step of the process, almost by
definition, cannot preserve all the information. The analog
measurement values are mapped to a discrete number of quantization
points. By the pigeonhole principle, it is impossible to represent an
infinite number of signals using a discrete number of values. Thus,
the goal of quantizer design is to exploit those values as efficiently
as possible to reduce the distortion on the signal.

One of the most popular methods for quantization is scalar
quantization. A scalar quantizer treats and quantizes each of the
signal measurements independently. This approach is particularly
appealing for its simplicity and its relatively good
performance. However, present approaches to scalar quantization do not
scale very well with the number of
measurements~\cite{bib:Thao94,bib:Thao96,bib:goyal98,Bou::2006::Quantization-and-erasures}. Specifically,
if the signal is oversampled, the redundancy of the samples is not
exploited effectively by the scalar quantizer. The trade-off between
the number of bits used to represent an oversampled signal and the
error in the representation does not scale well as oversampling
increases. In terms of the rate vs. distortion trade-off, it is
significantly more efficient to allocate representation bits such that
they produce refined scalar quantization with a critically sampled
representation as opposed to coarse scalar quantization with an
oversampled representation.

This trade-off can be reduced or eliminated using more sophisticated
or adaptive techniques such as vector quantization, Sigma-Delta
($\Sigma\Delta$)
quantization~\cite{bib:Thao96b,bib:Benedetto06,bib:BoufounosEurasip06},
or coding of level crossings~\cite{bib:Cvetkovic07}.  These methods
consider more than one sample in forming a quantized representation,
either using feedback during the quantization process or by grouping
and quantizing several samples together. These approaches improve the
rate vs. distortion trade-off significantly. The drawback is that each
of the measurements cannot be quantized independently, and they are
not appropriate when independent quantization of the coefficients is
necessary.

In this work we develop the basis for a measurement and scalar
quantization framework that significantly improves the rate-distortion
trade-off without requiring feedback or grouping of the
coefficients. Each measured coefficient is independently quantized
using a modified scalar quantizer with non-contiguous quantization
intervals. Using this modified quantizer we show that we can beat
existing lower bounds on the performance of oversampled scalar
quantization, which only consider quantizers with contiguous
quantization intervals~\cite{bib:Thao96,bib:Boufounos06thesis}.

The framework we present is universal in the sense that information
about the signal or the signal model is not necessary in the design of
the quantizer. In many ways, the quantization method is reminiscent of
information theoretic distributed coding results, such as the
celebrated Slepian-Wolf and Wyner-Ziv coding
methods~\cite{slepian1973noiseless,wyner_ziv_76}. While we only
analyze 1-bit scalar quantization, we discuss how the results can be
easily extended to multibit scalar quantization.

One of the key results we derive in this paper is the exponential
quantization error decay as a function of the oversampling rate. To
the best of our knowledge, it is the first example of a scalar
quantization scheme that achieves exponential error decay without
further coding or examination of the quantized samples. Thus, our
method is truly distributed in the sense that quantization and
transmission of each measurement can be performed independently of the
others. 

Our result has similar flavor with recent results in Compressive
Sensing, such as the Restricted Isometry Property (RIP) of random
matrices~\cite{bib:Candes05lp,bib:CandesCS06,CandesRIP,BarDavDeV::2008::A-Simple-Proof}.
Specifically, all our proofs are probabilistic and the results are
with overwhelming probability on the system parameters. The advantage
of our approach is that we do not impose a probabilistic model on the
acquired signal. Instead, the probabilistic model is on the
acquisition system, the properties of which are usually under the
control of the system designer.

The proof approach is inspired by the proof of the RIP of random
matrices in~\cite{BarDavDeV::2008::A-Simple-Proof}. Similarly
to~\cite{BarDavDeV::2008::A-Simple-Proof} we examine how the system
performs in distinguishing pairs of signals as a function of their
distance. We then extend the result on distinguishing a small ball
around each of the signals in the pair. By covering the set of signals
of interest with such balls we can extend the result to the whole
set. The number of balls required to cover the set and, by extension,
the Kolmogorov entropy of the set play a significant role in the
reconstruction performance. While Kolmogorov entropy is known to be
intimately related to the rate-distortion performance under vector
quantization, this is the first time is tied to the rate-distortion
performance under scalar quantization.

We assume a consistent reconstruction algorithm, i.e., an algorithm
that reconstructs a signal estimate that quantizes to the same
quantization values as the acquired
signal~\cite{bib:goyal98}. However, we do not discuss any practical
reconstruction algorithms in this paper. For any consistent
reconstruction algorithm it suffices to demonstrate that if the
reconstructed signal is consistent with the measurements, it cannot be
very different from the acquired signal. To do so, we need to examine
all the signals in the space we are interested in. Exploiting and
implementing these results with practical reconstruction algorithms is
a topic for future publications.

In the next section, which partly serves as a brief tutorial, we
provide an overview of the state of the art in scalar quantization. In
this overview we examine in detail the fundamental limitations of
current scalar quantization approaches and the reasons behind
them. This analysis suggests one way around the limitations, which we
examine in Sec~\ref{sec:analysis}.  In Sec.~\ref{sec:side_info} we
discuss the universality properties of our approach and we examine how
side-information on the signal can be incorporated in our framework to
improve quantization performance. In this spirit, we examine
Compressive Sensing and quantization of similar signals. Finally, we
discuss our results and conclude in Sec.~\ref{sec:discussion}.

\section{Overview of Scalar Quantization}
\label{sec:overview}
\subsection{Scalar Quantizer Operation}
A scalar quantizer operates directly on individual scalar signal
measurements without taking into account any information on the value
or the quantization level of nearby measurements. Specifically, the
generation of the $m^{th}$ quantized measurement from the quantized
signal $\vx\in\Reals^K$ is performed using
\begin{align}
  y_m&=\langle \vx, \phi_m \rangle +w_m \label{eq:measurement}\\
  q_m&=Q\left(\frac{y_m}{\Delta_m}\right), \label{eq:quantization}
\end{align}
where $\phi_m$ is the measurement vector and $w_m$ is the additive
dither used to produce a dithered scalar measurement $y_m$ which is
subsequently scaled by a precision parameter $\Delta_m$ and quantized
by the quantization function $Q(\cdot)$. The index $m=1,\ldots,M$,
where $M$ is the total number of quantized coefficients acquired. The
precision parameter is usually not explicit in the literature but is
incorporated as a design parameter of the quantization function
$Q(\cdot)$. We made it explicit in this overview in anticipation of
our development.

The measurement vectors can vary, depending on the problem at
hand. Typically they form a basis or an overcomplete frame for the
space in which the signal of interest
lies~\cite{bib:Christensen02,bib:goyal98,bib:Boufounos06thesis}. More
recently, Compressive Sensing demonstrated that it is possible to
undersample sparse signals and still be able to recover them using
incoherent measurement vectors, often randomly
generated~\cite{bib:Candes04A,bib:candes05st,bib:CandesCS06,bib:Candes04C,bib:Donoho04A}.
Random dither is sometimes added to the measurements to reduce certain
quantization artifacts and to ensure the quantization error has
tractable statistical properties. The dither is usually assumed to be
known and is taken into account in the reconstruction. If dither is
not used, $w_m=0$ for all $m$.

The quantization function $Q(\cdot)$ is typically a uniform quantizer,
such as the one shown in Fig.~\ref{fig:q_fun}(a) for a multi-bit
quantizer or in Fig.~\ref{fig:q_fun}(b) for a binary (1-bit)
quantizer. The number of bits required depends on the number of
quantization levels used by the quantizer. For example
Fig.~\ref{fig:q_fun}(a) depicts an 8-level, i.e. a $\log_2(8)=3$-bit
quantizer. The number of levels necessary, in turn, depends on the
dynamic range of the scaled measurements, i.e., the maximum and
minimum possible values, such that the quantizer does not overflow
significantly. A $B$-bit quantizer can represent of $2^B$ quantization
values, which determines the trade-off between accuracy and bit-rate.

The scaling performed by the precision parameter $\Delta_m$ controls
the trade-off between quantization accuracy and the number of
quantization bits. Larger $\Delta_m$ will cause a larger range of
measurement values to quantize to the same quantization level, thus
increasing the ambiguity and decreasing the precision of the
quantizer. Smaller values, on the other hand, increase the precision of
the quantizer but produce a larger dynamic range of values to be
quantized. Thus more quantization levels and, therefore, more bits
are necessary to avoid saturation.  Often non-uniform quantizers may
improve the quantization performance if there is prior knowledge about
the distribution of the measurements. These can be designed
heuristically, or using a design method such as the Lloyd-Max
algorithm~\cite{Lloyd82,Max60}. Recent work has also demonstrated that
overflow, if properly managed, can in certain cases be desirable and
effective in reducing the error due to
quantization~\cite{bib:LBDB_ACHA11,LasBouBar::2009::Finite-range-scalar}.
Even with these approaches, the fundamental accuracy vs. distortion
trade-off remains in some form.

A more compact, vectorized form of~\eqref{eq:measurement}
and~\eqref{eq:quantization} will often be more convenient in our
discussion
\begin{align}
  \vy&=\Phi\vx +\vw \label{eq:measurement_v}\\
  \vq&=\mQ\left(\Delta^{-1}\vy\right), \label{eq:quantization_v}
\end{align}
where \vy, \vq, and \vw\ are vectors containing the measurements, the
dither coefficients, and the quantized values, respectively, $\Delta$
is a diagonal matrix with the precision parameters $\Delta_m$ in its
diagonal, $\mQ(\cdot)$ is the scalar quantization applied
element-by-element on its input, and $\Phi$ is the $M\times K$
measurement matrix that contains the measurement vectors $\phi_m$ in
its rows.

\subsection{Reconstruction from Quantized Measurements}
A reconstruction algorithm, denoted $R(\cdot)$, uses the quantized
representation generated by the signal to produce a signal estimate
$\widehat{\vx}=R(\vq)$. The performance of the quantizer and the
reconstruction algorithm is measured in terms of the reconstruction
distortion, typically measured using the $\ell_2$ distance:
$d=\|\vx-\widehat{\vx}\|_2$. The goal of the quantizer and the
reconstruction algorithm is to minimize the average or the worst case
distortion given a probabilistic or a deterministic model of the
acquired signals.

The simplest reconstruction approach is to substitute the quantized
value in standard reconstruction approaches for unquantized
measurements. For example, if $\Phi$ forms a basis or a frame, we can
use linear reconstruction to compute
\begin{align*}
  \widehat{\vx}=\Phi^{\dagger}\left(\Delta\vq-\vw\right),
\end{align*}
where $(\cdot)^\dagger$ denotes the pseudoinverse (which is equal to
the inverse of $\Phi$ is a basis). Linear reconstruction using the
quantized values can be shown to be the optimal reconstruction method
if $\Phi$ is a basis. However, it is suboptimal in most other cases,
e.g., if $\Phi$ is an oversampled frame, or if Compressive Sensing
reconstruction algorithms are
used~\cite{bib:Thao96,bib:goyal98,bib:BoufounosDCC07}.

A better approach is to use consistent reconstruction, a
reconstruction method that enforces that the reconstructed signal
quantizes to the same value, i.e., satisfies the constraint
$\vq=\mQ\left(\Delta^{-1}\left(\Phi\widehat{\vx}
+\vw\right)\right)$. Consistent reconstruction was originally proposed
for oversampled frames in~\cite{bib:goyal98}, where it was shown to
outperform linear reconstruction. Subsequently consistent
reconstruction, or approximations of it, have been shown in various
scenarios to improve Compressive Sensing or other reconstruction from
quantized
measurements~\cite{bib:BoufounosCISS08,bib:LBDB_ACHA11,LasBouBar::2009::Finite-range-scalar,bib:Jacques09,bib:Zymnis09,bib:DaiPhaMil,bib:BoufounosAsilomar09,bib:BoufounosNonlinearICASSP10,vivekQuantFrame,bib:JLBB_BeSE}. It
is also straightforward to demonstrate that if $\Phi$ is a basis, the
simple linear reconstruction described above is also consistent.

\subsection{Reconstruction Rate and Distortion Performance}
The performance of scalar quantizers is typically measured by their
rate vs. distortion trade-off, i.e., how increasing the number of bits
used by the quantizer affects the distortion on the measurement signal
due to quantization. The distortion can be measured as worst-case
distortion, i.e.,
\begin{align*}
  d_{\mbox{wc}}=\max_\vx\left\|\vx-R\left(\mQ\left(\Delta^{-1}\left(\Phi\vx
+\vw\right)\right)\right)\right\|_2,
\end{align*}
or, if \vx\ is modeled as a random variable, average distortion,
\begin{align*}
  d_{\mbox{avg}}=E_{\vx}\left\{\left\|\vx-R\left(\mQ\left(\Delta^{-1}\left(\Phi\vx
+\vw\right)\right)\right)\right\|_2\right\},
\end{align*}
where $\widehat{\vx}=R\left(\mQ\left(\Delta^{-1}\left(\Phi\vx
+\vw\right)\right)\right)$ is the signal reconstructed from the
quantization of \vx.

In principle, under this sampling model, there are two ways to
increase the bit-rate and reduce the quantization distortion. The
first is to increase the number of bits used per quantized
coefficient. In terms of the description above, this is equivalent to
decreasing the precision parameter $\Delta_m$. For example, reducing
$\Delta_m$ by one half will double the quantization levels necessary
and, thus, increase the necessary bit-rate by 1 bit per
coefficient. On the other hand, it will decrease by 2 the ambiguity on
each quantized coefficient, and, thus, the reconstruction error. Using
this approach to increase the bit-rate, an exponential reduction in
the average error is possible as a function of the bit-rate
\begin{align}
  d=O(c^{r}), c\le 1,
  \label{eq:exp_rate}
\end{align}
where $r=MB$ is the total rate used to represent the signal at $M$
measurements and $B$ bits per measurement.

The second way is to increase the number of measurements at a fixed
number of bits per
coefficient. In~\cite{bib:Thao96,bib:Boufounos06thesis} it is shown
that the distortion (average or worst-case) cannot reduce at a rate
faster than linear with respect to the oversampling rate, which, at a
fixed number of bits per measurement, is proportional to the bit-rate;
i.e.,
\begin{align}
  d=\Omega(1/r),
  \label{eq:oversampling_bound}
\end{align}
much slower than the rate in~\eqref{eq:exp_rate}. It is further shown
in~\cite{bib:Thao96,bib:goyal98} that linear reconstruction does not
reach this lower bound, whereas consistent reconstruction approaches
do. Thus, the rate-distortion trade-off does not scale favorably when
increasing the number of measurements at a constant bit-rate per
measurement. A similar result can be shown for compressive acquisition
of sparse signals~\cite{bib:BoufounosDCC07}.

Despite the adverse trade-off, oversampling is an effective approach
to achieve
robustness~\cite{bib:Cvetkovic96,bib:goyal98,Goy::2001::Multiple-description,bib:Goyal01,bib:Puschel05,boufounos2008causal}
and it is desirable to improve this adverse trade-off. Approaches such
as Sigma-Delta quantization can be shown to improve the performance at
the expense of requiring feedback when computing the
coefficients. Even with Sigma-Delta quantization, the error decay
cannot become exponential in the oversampling rate~\cite{bib:Thao96b},
unless further coding is used~\cite{gunturk2003one}. This can be an
issue in applications where simplicity and reduced communication is
important, such as distributed sensor networks. It is, thus, desirable
to achieve scalar quantization where oversampling provides a favorable
rate vs. distortion trade-off, as presented in this paper.

\begin{figure*}
  \begin{minipage}{0.32\textwidth}
    \begin{center}
      \includegraphics[width=\textwidth]{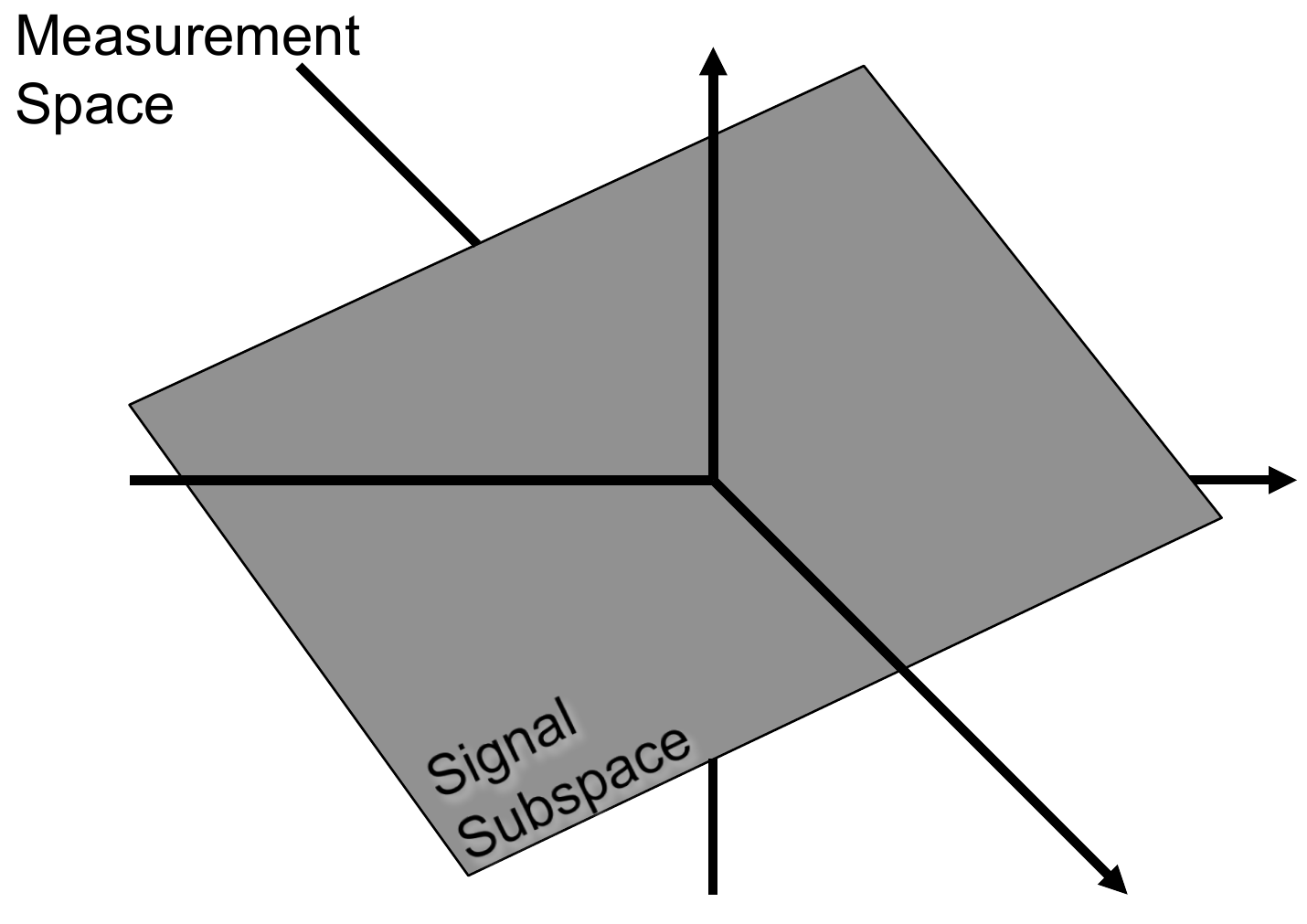}\\
      (a)
    \end{center}
  \end{minipage}
  \hfill
  \begin{minipage}{0.32\textwidth}
    \begin{center}
      \includegraphics[width=\textwidth]{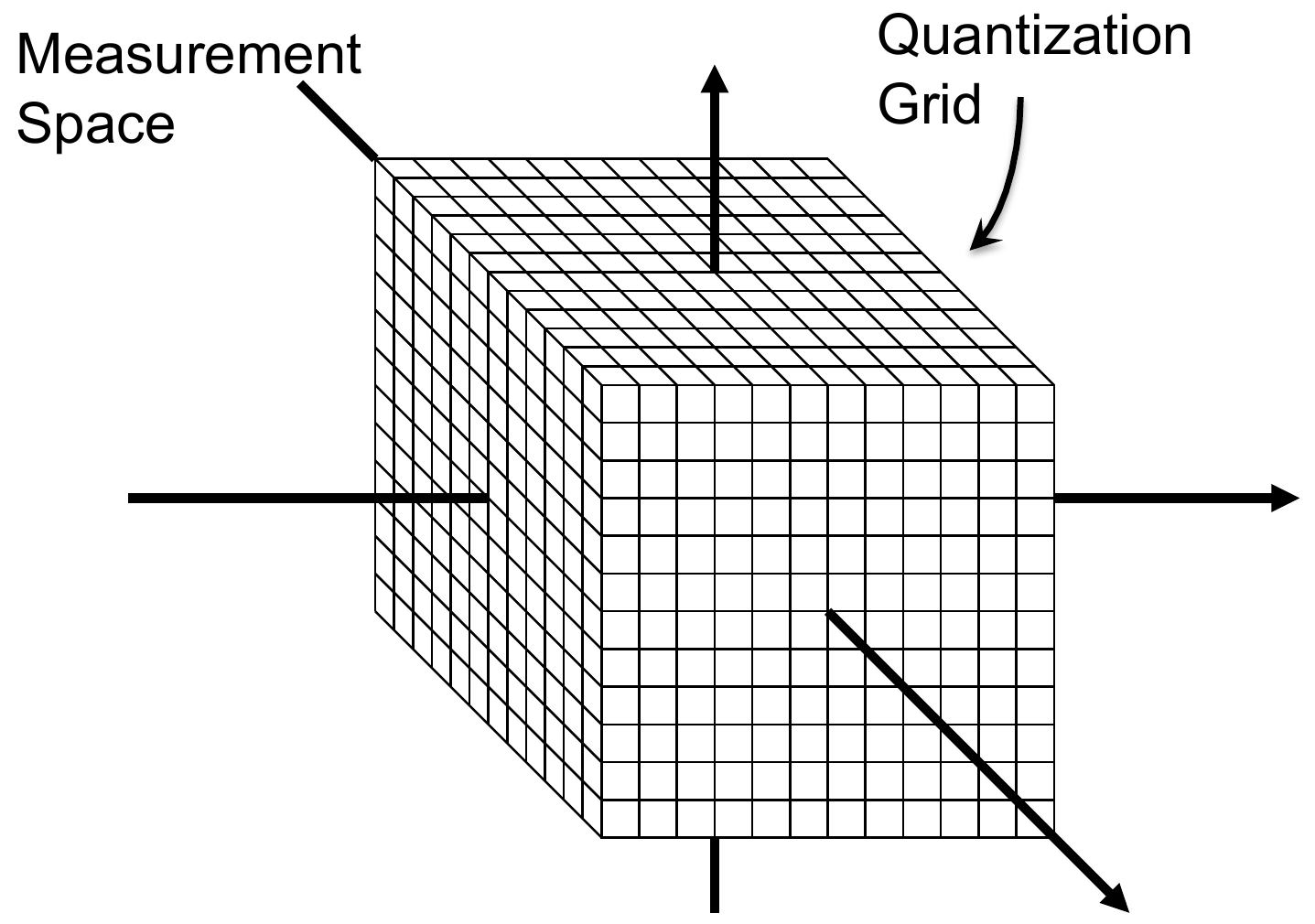}\\
      (b)
    \end{center}
  \end{minipage}
  \hfill
  \begin{minipage}{0.32\textwidth}
    \begin{center}
      \includegraphics[width=\textwidth]{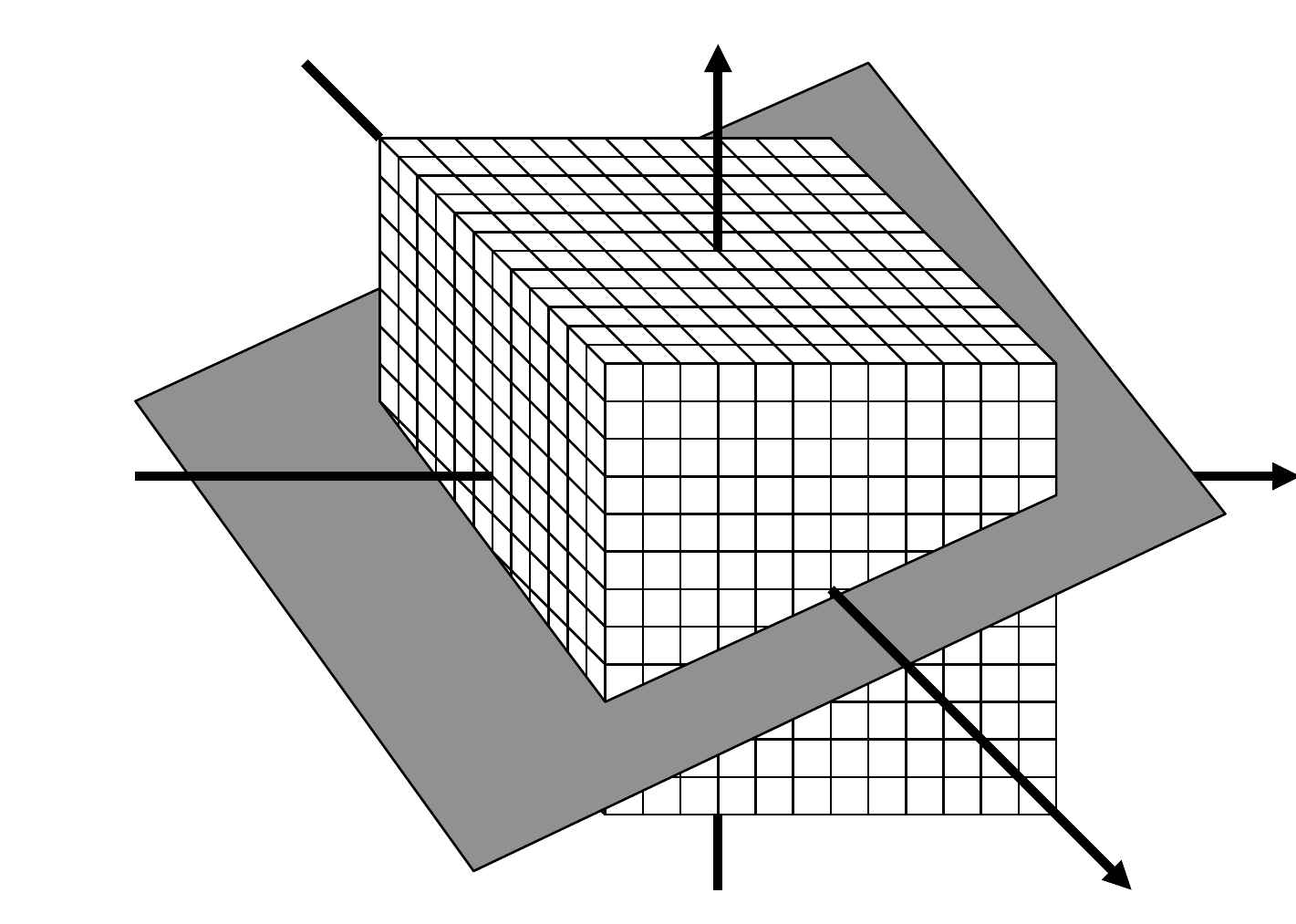}\\
      (c)      
    \end{center}
  \end{minipage}
  \caption{Oversampled Signals and Quantization. (a) Oversampled
    signals occupy only a small subspace in the measurement space. (b)
    The quantization grid quantizes all the measurement space. (c) The
    signal subspace intersects very few of the available quantization
    cells.}
  \label{fig:oversampled_signal}
\end{figure*}

The fundamental reason for this trade-off is the effective use of the
available quantization bits when oversampling. A linearly oversampled
$K$-dimensional signal occupies only a $K$-dimensional subspace (or
affine subspace, if dithering is used) in the $M$-dimensional
measurement space, as shown in
Fig.~\ref{fig:oversampled_signal}(a). On the other hand, the $2^{MB}$
bits used in the representation create quantization cells that equally
occupy the whole $M$-dimensional space, as shown in
Fig~\ref{fig:oversampled_signal}(b). The oversampled representation of
the signal will quantize to a particular quantization vector \vq\ only
if the $K$-dimensional plane intersects the corresponding quantization
cell. As evident in Fig~\ref{fig:oversampled_signal}(c), most of the
available quantization cells are not intersected by the plane, and
therefore most of the available quantization points \vq\ are not
used. Careful counting of the intersected cells provides the bound
in~\eqref{eq:oversampling_bound}~\cite{bib:Thao96,bib:Boufounos06thesis}.
The bound does not depend on the spacing of the quantization
intervals, or their size. A similar bound can be shown for a union of
$K$-dimensional subspaces, applicable in the case of Compressive
Sensing~\cite{bib:BoufounosDCC07,bib:JLBB_BeSE}.

\begin{figure*}
  \begin{minipage}{.22\textwidth}
    \begin{center}
      \includegraphics[width=\textwidth]{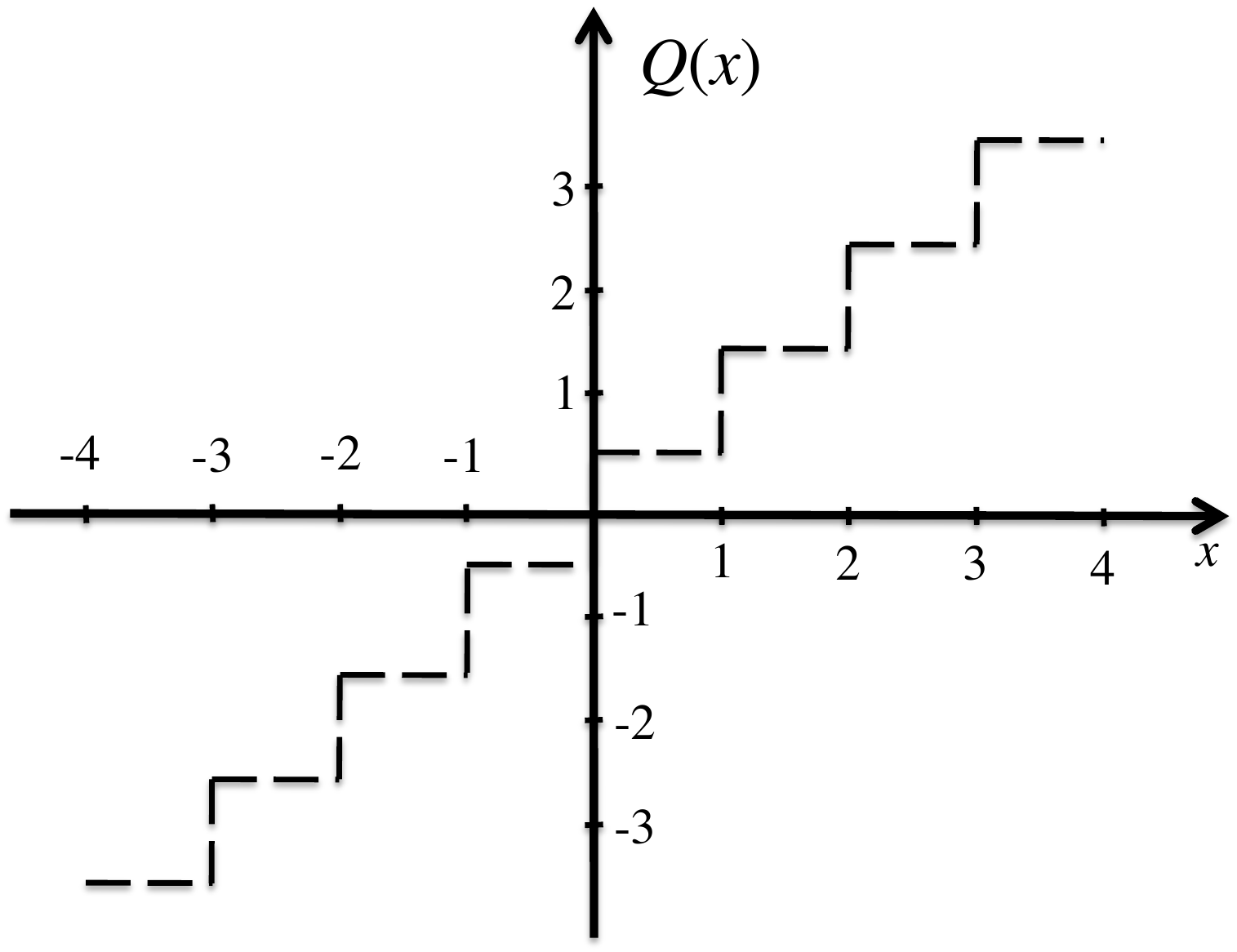}\\
    \end{center}
  \end{minipage}
  \hfill
  \begin{minipage}{.22\textwidth}
    \begin{center}
      \includegraphics[width=\textwidth]{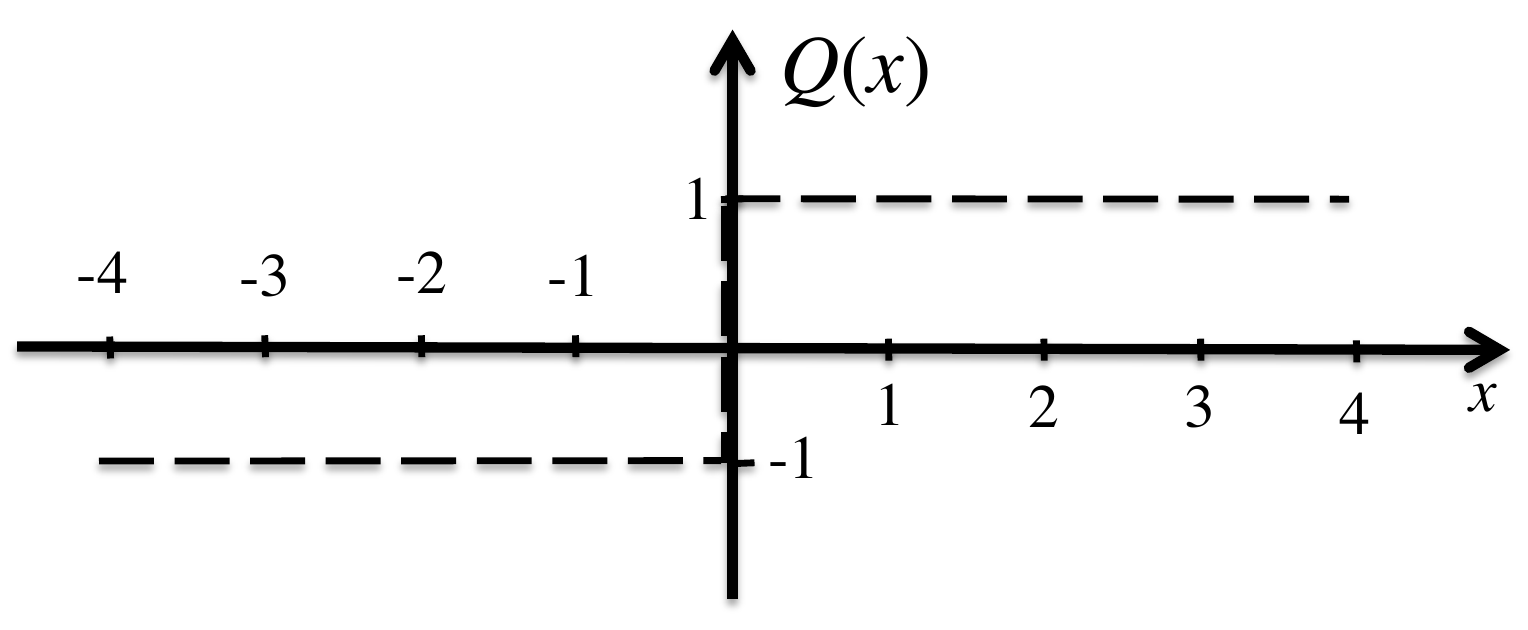}\\
    \end{center}
  \end{minipage}
  \hfill
  \begin{minipage}{.22\textwidth}
    \begin{center}
      \includegraphics[width=\textwidth]{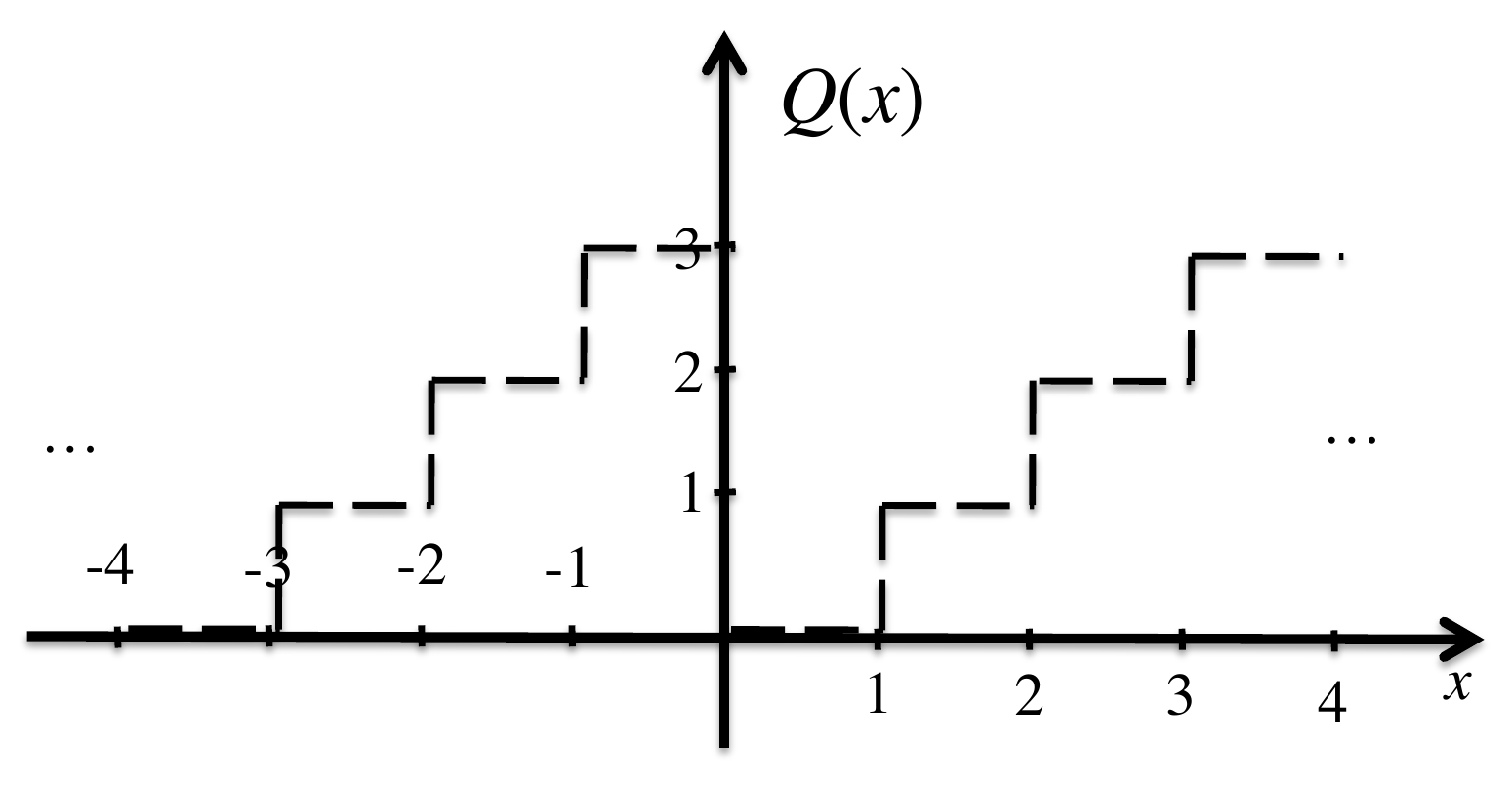}\\
    \end{center}
  \end{minipage}
  \hfill
  \begin{minipage}{.22\textwidth}
    \begin{center}
      \includegraphics[width=\textwidth]{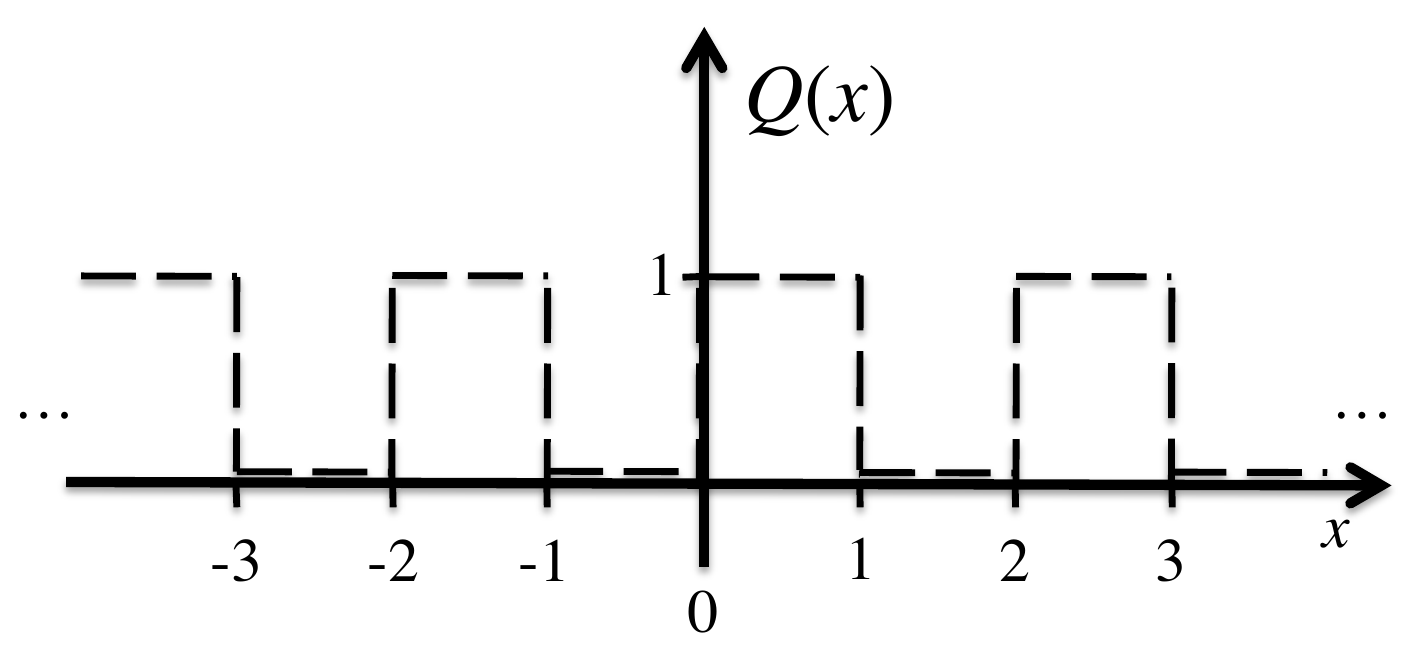}\\
    \end{center}
  \end{minipage}
  \\
  \begin{minipage}{.22\textwidth}
    \centerline{(a)}
  \end{minipage}
  \hfill
  \begin{minipage}{.22\textwidth}
    \centerline{(b)}
  \end{minipage}
  \hfill
  \begin{minipage}{.22\textwidth}
    \centerline{(c)}
  \end{minipage}
  \hfill
  \begin{minipage}{.22\textwidth}
    \centerline{(d)}
  \end{minipage}
  \caption{Examples of Quantization Functions. Typical (a) multibit
    and (b) binary (1-bit) quantization functions used in scalar
    quantization. Proposed (c) multibit and (d) binary quantization
    functions, used in this work.}
  \label{fig:q_fun}
\end{figure*}

To overcome the adverse trade-off, a scalar quantizer should be able
to use most of the $2^{MB}$ available quantization vectors, i.e.,
intersect most of the available quantization cells. Note that
no-matter how we choose the quantization intervals, the shape of the
quantization cells is rectangular and aligned with the axes. Thus,
improving the trade-off requires a strategy other than changing the
shape and positioning of the quantization cells. The approach we use
in this paper is to make the quantization cells non-continuous by
making the quantization function non monotonic, as shown in
Figs.~\ref{fig:q_fun}(c) and~\ref{fig:q_fun}(d). This is, in many
ways, similar to the binning of quantization cells explored
experimentally in~\cite{bib:Pai06}. The advantage of our approach is
that it facilitates theoretical analysis and can scale down to even
one bit per measurement. In the remainder of this paper we demonstrate
that our proposed approach achieves, with very high probability,
exponential decay in the worst-case quantization error as a function
of the oversampling rate, and, consequently, the bit-rate.

\section{Rate-efficient Scalar Quantization}
\label{sec:analysis}
\subsection{Overview}
Our approach uses the scalar quantizer described
in~\eqref{eq:measurement} and~\eqref{eq:quantization} with the
quantization function in Figs.~\ref{fig:q_fun}(c)
and~\ref{fig:q_fun}(d). The quantization function is explicitly
designed to be non-monotonic, such that non-contiguous quantization
regions quantize to the same quantization value. This allows the
subspace defined by the measurements to intersect the majority of the
available quantization cells which, in turn, ensures efficient use of
the available bit-rate. Although we do not describe a specific
reconstruction algorithm, we assume that the reconstruction algorithm
produces a signal consistent with the measurements, in addition to
imposing a signal model or other application-specific requirements.

Our end goal is to determine an upper bound for the probability that
there exist two signals \vx\ and $\vx'$ with distance greater than $d$
that quantize to the same quantization vector given the number of
measurements $M$. If no such pair exists, then any consistent
reconstruction algorithm will reconstruct a signal that has distance
from the acquired signal at most $d$. We wish to demonstrate that this
probability vanishes very fast as the number of measurements
increases. Furthermore, we wish to show that for a fixed probability
of such a signal pair existing, the distance to guarantee such
probability decreases exponentially with the number of
measurements. An important feature of our development is that the
probability of success is on the acquisition system randomization,
which we control, and not on any probabilistic model for the signals
acquired.

To achieve our goal we first consider a single measurement on a pair
of signals \vx, and $\vx'$ with distance $d=\|\vx-\vx'\|_2$, and
analyze the probability a single measurement of the two signals
quantizes to the same quantization value for both. Our result is
summarized in Lemma~\ref{th:single_pair}.

\begin{lemma}
  Consider signals \vx, and $\vx'$ with $d=\|\vx-\vx'\|_2$ and the
  quantized measurement
  function \[q=Q\left(\frac{\langle\vx,\phi\rangle+w}{\Delta}\right),\]
  where $Q(x)=\lceil x \rceil~\mod~2$, $\phi_m\in\Reals^K$ contains
  i.i.d. elements drawn from a normal distribution with mean 0 and
  variance $\sigma^2$, and $w_k$ is i.i.d., uniformly distributed in
  $[0,\Delta]$.

  The probability
  that the two signals produce equal quantized measurements is
  \[
    P(\vx,\vx'\mathrm{~consistent}|d) = \frac{1}{2}
    +\sum_{i=0}^{+\infty}\frac{e^{-\left(\frac{\pi(2i+1)\sigma d}
        {\sqrt{2}\Delta}\right)^2}}{\left(\pi(i+1/2)\right)^2}\le
    \frac{1}{2}+\frac{1}{2}e^{-\left(\frac{\pi\sigma
        d}{\sqrt{2}\Delta}\right)^2}.
    \]
  \label{th:single_pair}
\end{lemma}

\vspace{0.1in}

We prove this lemma in Sec.~\ref{sec:signal_pairs}.

Next, in Sec.~\ref{sec:epsilon_balls}, we consider a single
measurement on two $\epsilon$-balls, \ball{\epsilon}{\vx}\ and
\ball{\epsilon}{\vx'}, centered at \vx\ and $\vx'$, i.e., on all the
signals of distance less than $\epsilon$ from \vx\ and $\vx'$. Using
Lemma~\ref{th:single_pair}, we lower-bound the probability that no
signal in \ball{\epsilon}{\vx}\ is consistent with any signal in
\ball{\epsilon}{\vx'}. This leads to Lemma~\ref{th:single_balls}.

\begin{lemma}
  Consider signals \vx, and $\vx'$ with $d=\|\vx-\vx'\|_2$, the
  $\epsilon$-balls \ball{\epsilon}{\vx}\ and
  \ball{\epsilon}{\vx'}\ and the quantized measurement function in
  Lemma~\ref{th:single_pair}.

  The probability that no signal in \ball{\epsilon}{\vx}\ produces
  equal quantized measurement with any signal in
  \ball{\epsilon}{\vx'}\ (i.e. the probability that the two balls
  produce inconsistent measurements) is lower bounded by
  \[
    P(\ball{\epsilon}{\vx},\ball{\epsilon}{\vx'}\mathrm{~inconsistent}|d)
    \ge 1- \left(P(\vx,\vx'\mathrm{~consistent}|d)+
    \frac{2c_p\epsilon}{\Delta}+\gamma\left(\frac{K}{2},\left(\frac{c_p}{2\sigma}\right)^2\right)\right),
    \]  \label{th:single_balls}
for any choice of $c_P\le \Delta/2\epsilon$, where $\gamma(s,x)$ is
the regularized upper incomplete gamma function.
\end{lemma}

\vspace{0.1in}

Finally we construct a covering of the signal space under
consideration using $\epsilon$-balls. We consider all pairs of
$\epsilon$-balls in this covering and using
Lemma~\ref{th:single_balls} we lower bound the probability than no
pair of signals with distance greater than $d$ produces consistent
measurements. This produces the main result of this work, proven in
Sec.~\ref{sec:main_result}.

\begin{theorem}
  Consider the set of signals
  \[\sS=\left\{\left.\vx\in\Reals^K\right|\|\vx\|_2\le
  1\right\}\] and the measurement system
  \[q_m=Q\left(\frac{\langle\vx,\phi_m\rangle+w_m}{\Delta}\right),
  ~m=1,\ldots,M,\] where $Q(x)=\lceil x \rceil~\mod~2$,
  $\phi_m\in\Reals^K$ contains i.i.d. elements drawn from a standard
  normal distribution, $w_k$ is i.i.d., uniformly distributed in
  $[0,\Delta]$.

  For any $c_r>1/2$, arbitrarily close to $1/2$, there exists a
  constant $c_o$ and a choice of $\Delta$ proportional to $d$ such
  that with probability greater than
  \[P\ge1-\left(\frac{c_o\sqrt{K}}{d}\right)^{2K}\left(c_r\right)^M\]
  the following holds for all $\vx,\vx'\in\sS$
  \[\|\vx-\vx'\|_2\ge d~\Rightarrow~\vq\ne\vq',\]
  where $\vq$ and $\vq'$ are the vectors containing the quantized
  measurements of $\vx$ and $\vx'$, respectively.
  \label{th:main}
\end{theorem}

\vspace{0.1in}

The theorem trades-off the leading term
$\left(\frac{c_o\sqrt{K}}{d}\right)^{2K}$ with how close to 1/2 is
$c_r$, i.e., how fast the probability in the statement approaches 1 as
a function of the number of measurements. Using an example, we also
make this result concrete and show that for $K>8$, we can achieve
$c_o=60,~c_r=3/4$.

Our results do not assume a probabilistic model on the signal. Instead
they are similar in nature to many probabilistic results in
Compressive
Sensing~\cite{bib:Candes04A,bib:Candes04C,bib:Candes05lp,bib:candes05st,bib:CandesCS06,CandesRIP,BarDavDeV::2008::A-Simple-Proof}.
With overwhelming probability the system works on all signals
presented to it.  It is also important to note that the results are
not asymptotic, but hold for finite $K$ and $M$. Further note that the
alternative is not that the system provides incorrect result, only
that we cannot guarantee that it will provide correct results.  Thus,
we fix the probability that we cannot guarantee the results to $P_0$
and demonstrate the desired exponential decay of the error.

\begin{corollary}
  Consider the set of signals, the measurement system and the
  consistent reconstruction process implied by
  Thm.~\ref{th:main}. With probability greater than
  \[P\ge 1-P_0,\]
  the following holds for all $\vx,\vx'\in\sS$
  \[\|\vx-\vx'\|_2\ge\frac{c_o\sqrt{K}}{P_0^{\frac{1}{2k}}}
  \left(c_r\right)^{\frac{M}{2K}} 
  \Rightarrow \vq\ne\vq'\]
  \label{th:exp_decay}
\end{corollary}

\vspace{0.1in}

The corollary makes explicit the exponential decay of the worst-case
error as a function both of the number of measurements $M$ and the
number of bits used. This means that the worst-case error decays
significantly faster than the linear decay demonstrated with classical
quantization of oversampled frames\cite{bib:Thao94,bib:goyal98} and
defeats the lower bound in~\cite{bib:Thao96}. Furthermore, we achieve
that rate by quantizing each coefficient independently, unlike
existing approaches~\cite{bib:Cvetkovic07,gunturk2003one}. Since this
is a probabilistic result on the system probability space, it further
implies that a system that satisfies the desired exponential decay
property exists.  

One of the drawbacks of this approach is that it requires the
quantizer to be designed in advance with the target distortion in
mind, i.e., the choice of the scaling parameter $\Delta$ of the
quantizer affects the distortion. This might be an issue if the target
accuracy and oversampling rate is not known at the quantizer design
stage, but, for example, needs to be estimated from the measurements
adaptively during measurement time. This drawback, as well as one way
to overcome it, is discussed further in Sec.~\ref{sec:discussion}.

The remainder of this section presents the above results in sequence.

\subsection{Quantized Measurement of Signal Pairs}
\label{sec:signal_pairs}
We first consider two signals $\vx$ and $\vx'$ with $\ell_2$ distance
$d=\|\vx-\vx'\|_2$. We analyze the probability that a single quantized
measurement of the two signals produces the same bit values, i.e., is
consistent for the two signals. Since we only discuss the measurement
of a single bit, we omit the subscript $m$ from~\eqref{eq:measurement}
and~\eqref{eq:quantization} to simplify the notation in the remainder
of this section. The analysis does not depend on it. We use $q$ and
$q'$ to denote a single quantized measurement of $\vx$ and $\vx'$,
respectively.

\begin{figure*}
  \begin{minipage}{.5\textwidth}
    \begin{center}
      \includegraphics[width=\textwidth]{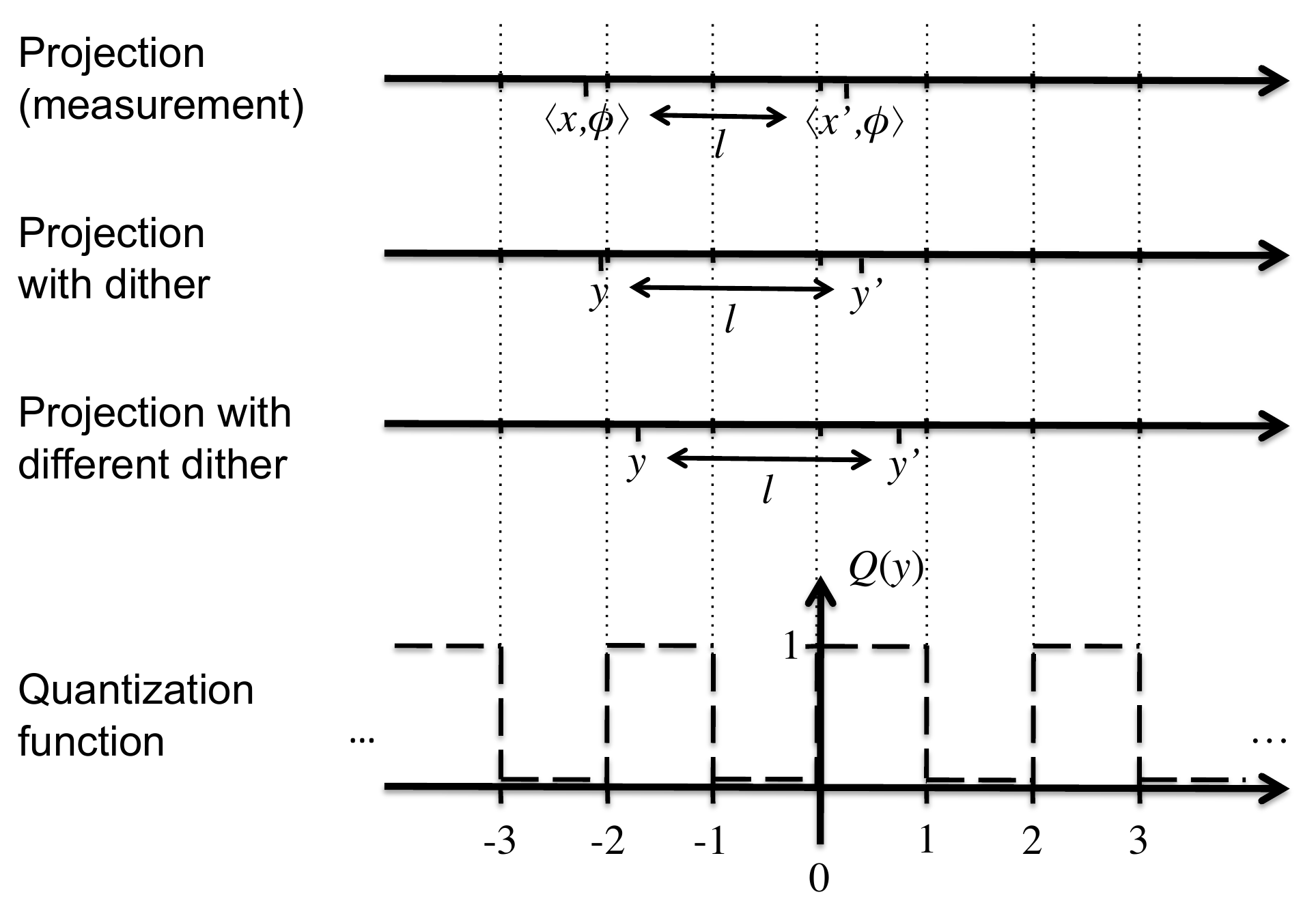}
      (a)
    \end{center}
  \end{minipage}
  \hfill
  \begin{minipage}{.4\textwidth}
    \begin{center}
      \includegraphics[width=\textwidth]{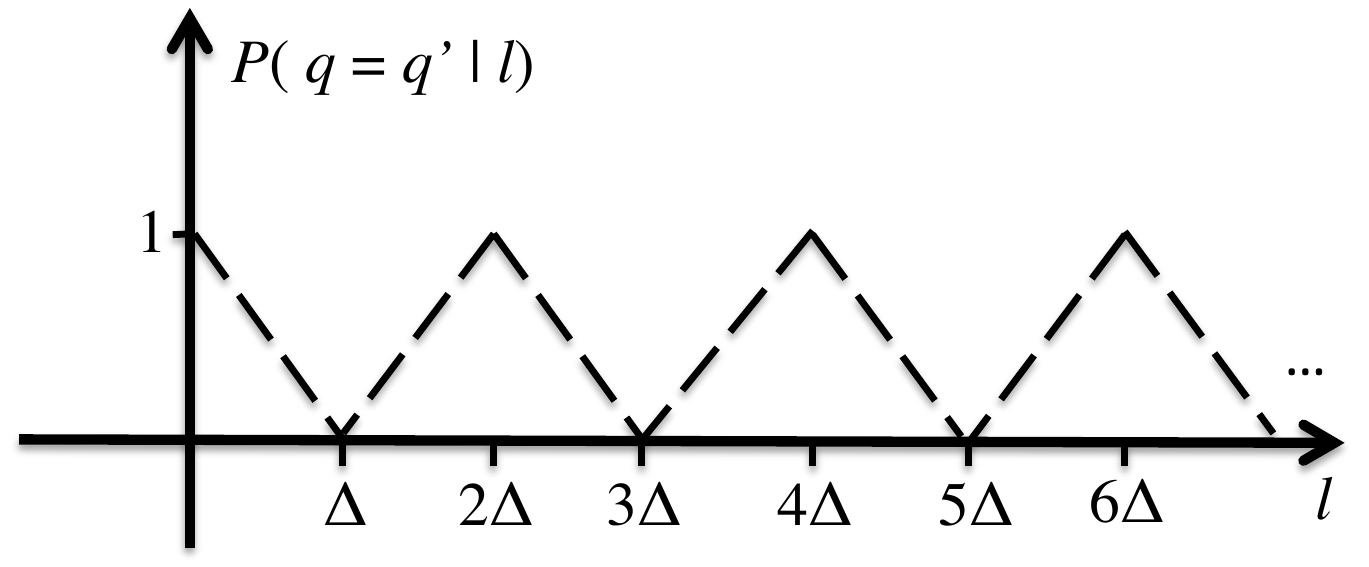}\\
      (b)\\
      
      \vspace{0.2in}

      \includegraphics[width=\textwidth]{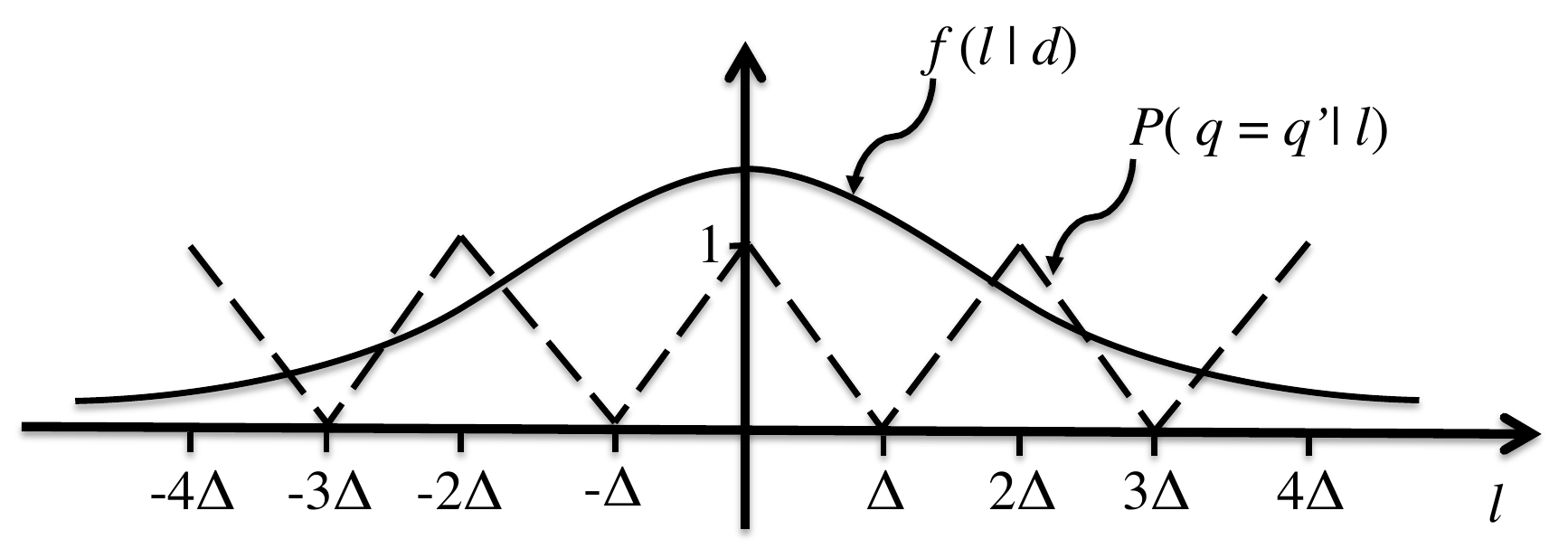}\\
      (c)
    \end{center}
  \end{minipage}
  \caption{Analysis of the probility of consistency for a single
    bit. (a) Dithering makes this probability depend only on the
    distance between the two signals. (b) Probability of consistency
    as a function of the projection length. (c) The two components
    affecting the overall probability of consistency.}
  \label{fig:consistent}
\end{figure*}

We first consider the desired probability conditional on the projected
distance $l$, i.e., the distance between the measurements of the
signals
\begin{align}
  l&\equiv |y-y'|=|\langle \vx,\phi\rangle+w-(\langle \vx',\phi\rangle+w)|\nonumber\\
  \Rightarrow l&=|\langle \vx-\vx',\phi\rangle|
  \label{eq:projected_distance}
\end{align}
The addition of dither makes the probability the two signals quantize
to consistent bits depend only on the distance $l$ and not on the
individual values $y$, and $y',$ as demonstrated in
Fig.~\ref{fig:consistent}(a). In the top part of the figure an example
measurement is depicted. Depending on the amount of dither, the two
measurements can quantize to different values (as shown in the second
line of the plot) or to the same values (as shown in the third
line). Since the dither is uniform in $[0,\Delta]$, the probability
the two bits are consistent given $l$ equals
\begin{align}
  P(q=q'|l)=\left\{
  \begin{array}{rcr}
    1-\frac{l\mod\Delta}{\Delta}=1+2i-\frac{l}{\Delta},&
    \mathrm{if}&2i\Delta\le l \le (2i+1)\Delta\\
    \frac{l\mod\Delta}{\Delta}=\frac{l}{\Delta}-(2i+1),&
    \mathrm{if}&(2i+1)\Delta\le l \le 2(i+1)\Delta,
  \end{array}
  \right.
  \label{eq:projected_cons}
\end{align}
for some integer $i$, which is plotted in
Fig.~\ref{fig:consistent}(b).

Furthermore, from~\eqref{eq:projected_distance} and the distribution
of $\phi$, it follows that $l$ is distributed as the magnitude of the
normal distribution with variance $(\sigma d)^2$
\begin{align*}
  f(l|d)=\sqrt{\frac{2}{\pi}}\frac{e^{-\left(\frac{l}{2\sigma
        d}\right)^2}}{\sigma d},~l\ge 0.
\end{align*}
Thus, the two quantization bits are the same given the distance of the
signals $d$ with probability
\begin{align}
  P(q=q'|d)=\int_{l\ge 0}P(q=q'|l)\cdot f(l|d)\mathrm{d}l.
  \label{eq:pr_consistent}
\end{align}

In order to evaluate the integral, we make it symmetric around zero by
mirroring it and dividing it by two. The two components of the
expanded integral are shown in Fig.~\ref{fig:consistent}(c). These are a
periodic triangle function with height 1 and width $2\Delta$ and a
Normal distribution function with variance $(\sigma d)^2$. 

Using Parseval's theorem, we can express that integral in the Fourier
domain (with respect to $l$). Noting that the periodic triangle
function can also be represented as a convolution of a single triangle
function with an impulse train, we obtain:
\begin{align*}
  P(q=q'|d)&=\int_{\xi}\stackrel{P(q=q'| l)}{\overbrace{\frac{1}{2}
      \sum_{i=-\infty}^{+\infty}\sinc^2\left(\frac{i}{2}\right)
      \delta\left(\xi-\frac{i}{2\Delta}\right)}}\stackrel{f(l|d)}
  {\overbrace{e^{-2(\pi\xi\sigma
        d)^2}}}\mathrm{d}\xi\\ &=\sum_{i=-\infty}^{+\infty}
  \frac{\sinc^2\left(\frac{i}{2}\right)}{2} e^{-\left(\frac{\pi
      i\sigma d}{\sqrt{2}\Delta}\right)^2}
\end{align*}
where $\sinc(x)\equiv\frac{\sin(\pi x)}{\pi x}$, and $\xi$ is the
frequency with respect to $l$. Since $\sinc(x)=0$ if $x$ is a non-zero
integer, $\sin^2(\pi x/2)=1$ if $x$ is an odd integer, and
$\sinc(0)=1$,
\begin{align}
  P(q=q'|d) &= \frac{1}{2}
  +\sum_{i=0}^{+\infty}\frac{e^{-\left(\frac{\pi(2i+1)\sigma d}
      {\sqrt{2}\Delta}\right)^2}}{\left(\pi(i+1/2)\right)^2},
  \label{eq:pr_cons_sum}
\end{align}
which proves the equality in Lemma~\ref{th:single_pair}.

\begin{figure}
  \centerline{\includegraphics[width=0.4\textwidth]{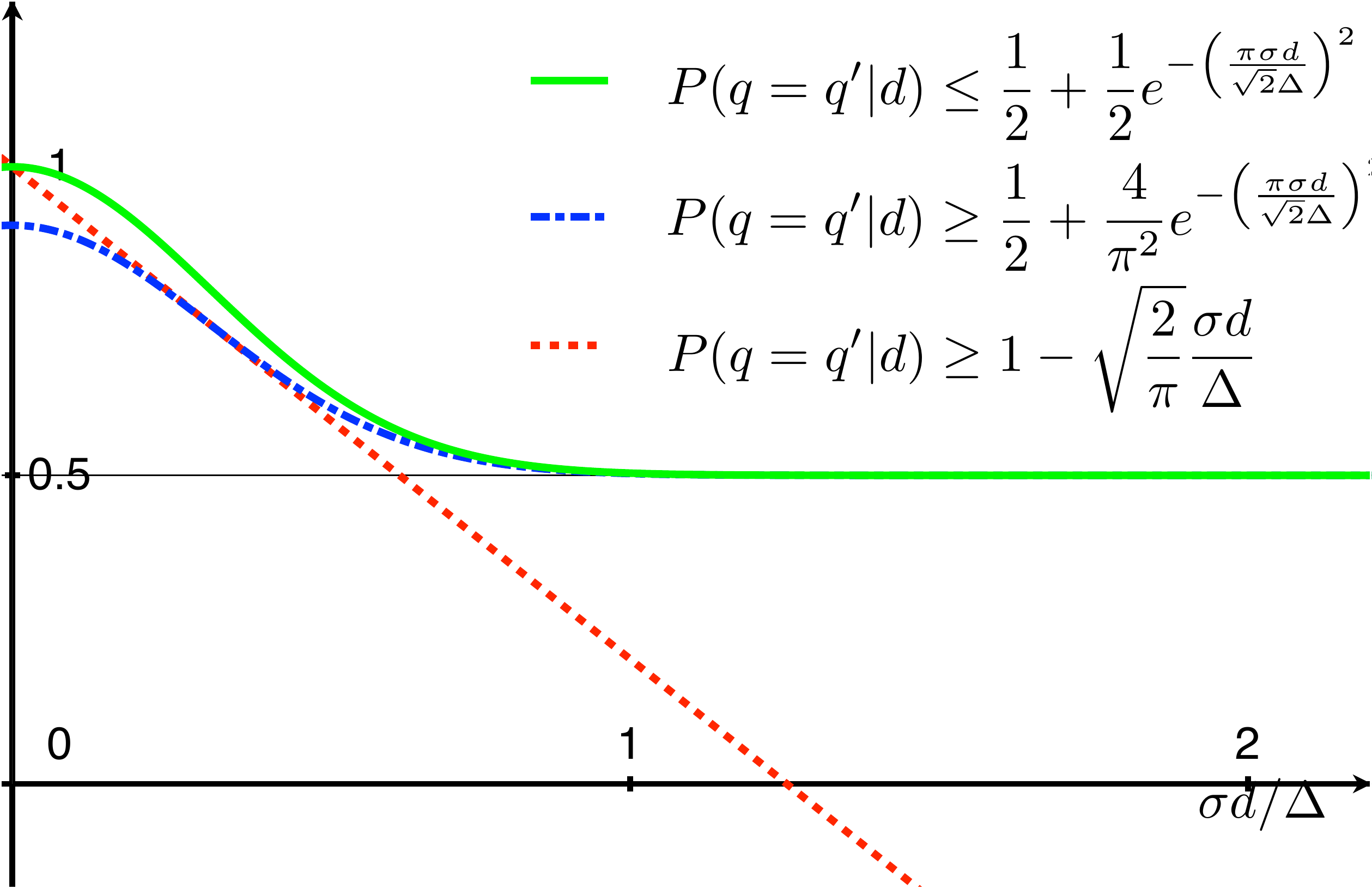}}
  \caption{Upper and lower bounds for the probability two different
    signals have consistent quantization bits}
  \label{fig:simplebounds}
\end{figure}

A very good lower bound for~\eqref{eq:pr_cons_sum} can be derived
using the first term of the summation:
\begin{align*}
  P(q=q'|d) &\ge \frac{1}{2}+\frac{4}{\pi^2}e^{-\left(\frac{\pi\sigma
        d}{\sqrt{2}\Delta}\right)^2}
\end{align*}
An alternative lower bound can also be derived by explicitly
integrating~\eqref{eq:pr_consistent} up to $l\le\Delta$:
\begin{align*}
  P(q=q'|d) &\ge 1-\sqrt{\frac{2}{\pi}}\frac{\sigma d}{\Delta}
\end{align*}
An upper bound can be derived using $e^{-\left(\frac{\pi(2i+1)\sigma
    d} {\sqrt{2}\Delta}\right)^2}\le e^{-\left(\frac{\pi\sigma d}
  {\sqrt{2}\Delta}\right)^2}$ in the summation in
\eqref{eq:pr_cons_sum}, and noting that $P(q=q'|d=0)=1$.
\begin{align*}
  P(q=q'|d) &\le \frac{1}{2}+\frac{1}{2}e^{-\left(\frac{\pi\sigma
        d}{\sqrt{2}\Delta}\right)^2},
\end{align*}
which proves the inequality and concludes the proof of
Lemma~\ref{th:single_pair}.  Fig.~\ref{fig:simplebounds} plots these
bounds and illustrates their tightness. A bound for the probability of
inconsistency can be determined from the bounds above using $P(q\ne
q'|d) = 1-P(q=q'|d)$.

Using these results it is possible to further analyse the performance
of this method on finite sets of signals. However, for many signal
processing applications it is desirable to analyse infinite sets of
signals, such as signal spaces. To facilitate this analysis the next
section examines how the system behaves on pairs of $\epsilon$-balls
in the signal space.

\subsection{Consistency of $\epsilon$-Balls}
\label{sec:epsilon_balls}
In this section we examine the performance on pairs of sets of
signals. Specifically, the sets we consider are $\epsilon$-balls in
$\Reals^K$ with radius $\epsilon$ and centered at \vx, defined as
\begin{align*}
  \ball{\epsilon}{\vx}=
  \left\{\left.\vs\in\Reals^K\right|\|\vs-\vx\|_2\le\epsilon\right\}.
\end{align*}
We examine balls $\ball{\epsilon}{\vx}$ and $\ball{\epsilon}{\vx'}$
around two signals \vx\ and $\vx'$ with distance $d=\|\vx-\vx'\|_2$,
as above. We desire to lower bound the probability that the quantized
measurements of all the signals in $\ball{\epsilon}{\vx}$ are
consistent with each other, and inconsistent with the ones from all
the signals in $\ball{\epsilon}{\vx'}$.

To determine the lower bound, we examine how the measurement vector
$\phi$ affects the $\epsilon$-balls. It is straightforward to show
that the measurement projects the \ball{\epsilon}{\vx}\ to an interval
in $\Reals$ of length at most $2\epsilon\|\phi\|_2$, centered at
$\langle \vx,\phi\rangle$. The length of the interval affects the
probability that the measurements of all the signals in
\ball{\epsilon}{\vx}\ quantize consistently. To guarantee consistency
we bound the length of this interval to be smaller than $2c_p\epsilon$,
i.e., we require that $\|\phi\|_2\le c_p$. This fails with probability
\begin{align*}
  P(\|\phi\|_2\ge
  c_p)&=\gamma\left(\frac{K}{2},\left(\frac{c_p}{2\sigma}\right)^2\right),
\end{align*}
where $\gamma(s,x)$ is the regularized upper incomplete gamma
function, and
$\gamma\left(\frac{K}{2},\left(\frac{x}{2}\right)^2\right)$ is the
tail integral of the $\chi$ distribution with $K$ degrees of freedom
(i.e., the distribution of the norm of a $K$-dimensional standard
Gaussian vector). To ensure that all the signals in the
$\epsilon$-ball can quantize to the same bit value with non-zero
probability we pick $c_p$ such that $2c_p\epsilon<\Delta$.

\begin{figure}
  \begin{minipage}{0.5\textwidth}
    \centerline{\includegraphics[width=\textwidth]{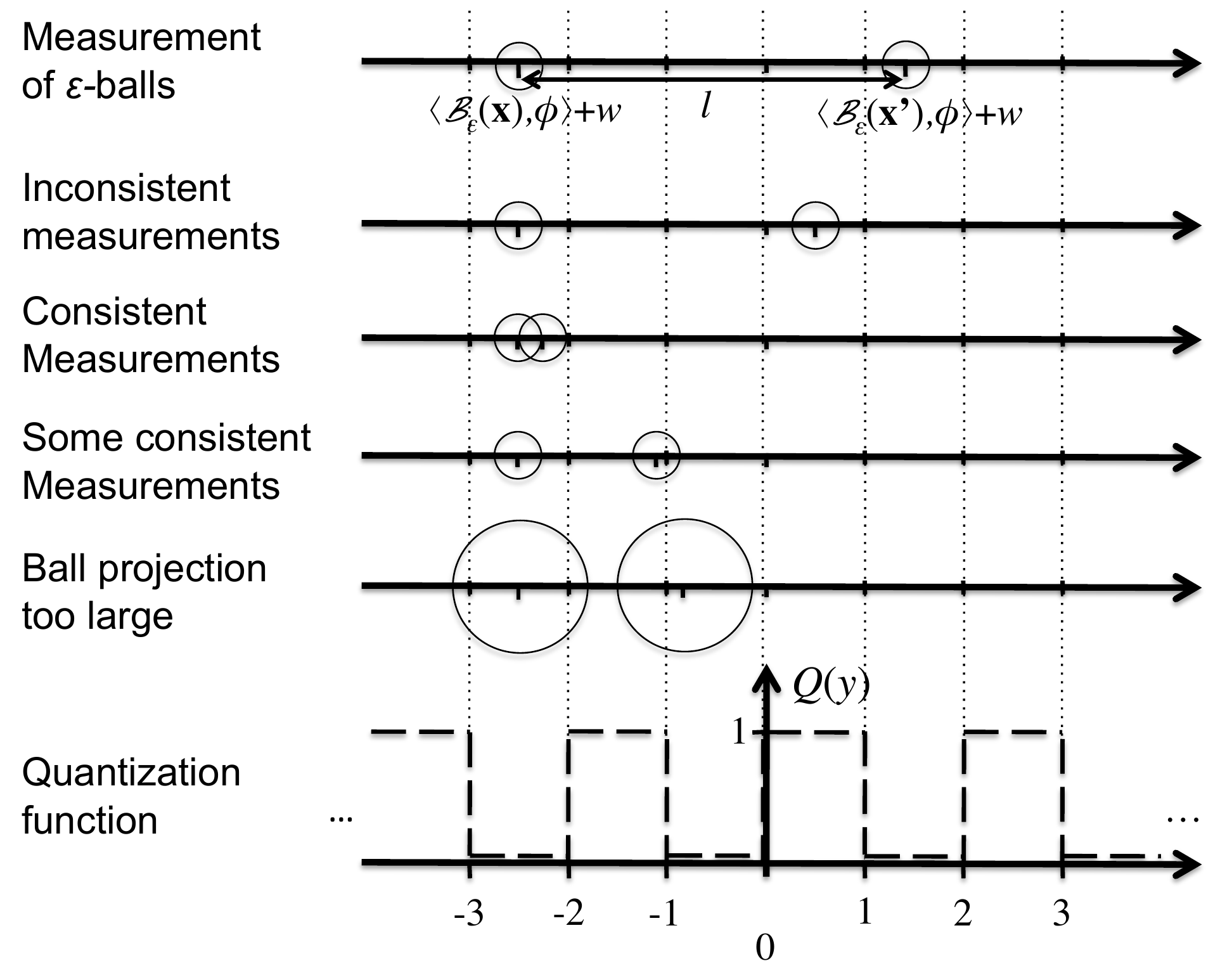}}
  \end{minipage}
  \hfill
  \begin{minipage}{0.4\textwidth}
  \centerline{\includegraphics[width=\textwidth]{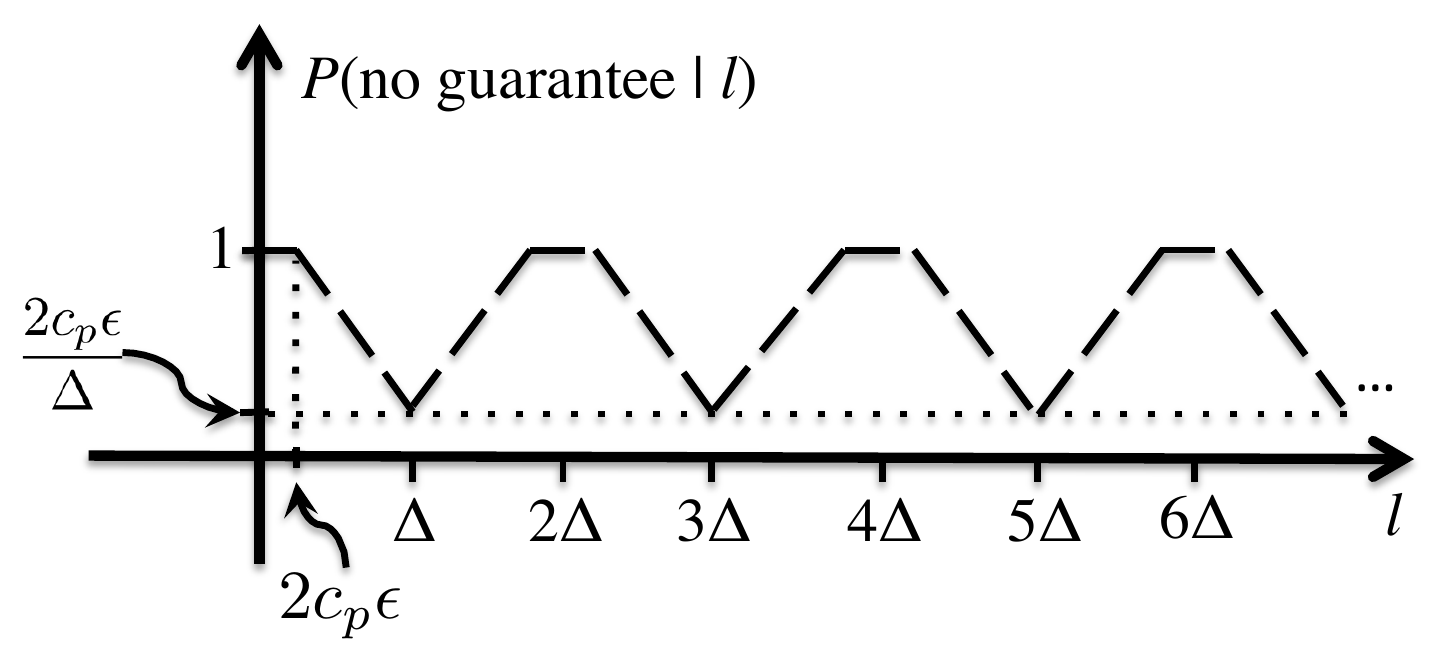}}
  \end{minipage}\\

  \begin{minipage}{0.5\textwidth}
  \centerline{(a)}
  \end{minipage}
  \hfill
  \begin{minipage}{0.4\textwidth}
  \centerline{(b)}
  \end{minipage}
  \caption{Measurement quantization of $\epsilon$-balls. (a) Ball
    measurement and consistency behavior. (b) Probability of no
    consistency guarantee given the projection
    length,~\eqref{eq:no_guarantee}.}
  \label{fig:ball_quantization}
\end{figure}

Under this restriction, the two balls will produce inconsistent
measurements only if the two intervals they project onto are located
completely within two quantization intervals with different
quantization values. Thus we cannot guarantee consistency within the
ball if the ball projection is on the boundary of a quantization
threshold, and we cannot guarantee inconsistency between the balls if
parts of the projections of the two balls quantize to the same
bit. Figure~\ref{fig:ball_quantization}(b) demonstrates the
quantization of $\epsilon$-balls, and examines when all the elements
of the two balls quantize inconsistently.

Assuming that the width of the ball projections is bounded, as
described above, then we can characterize the probability that the
ball centers will project on the quantization grid in a way that all
signals within one ball quantize to the same one quantization value,
and all the signals from the other ball quantize to the other. This is
the probability that we can guarantee that all measurements from the
signals in one ball are inconsistent with all the signals from the
other ball. We desire to upper bound the probability that we fail to
guarantee this inconsistency.

Using, as before, $l$ to denote the projected distance between the two
centers of the balls, we cannot guarantee inconsistency if
$|l-2i\Delta|\le 2c_p\epsilon$ for some $i$. In this case the balls
are guaranteed to intersect modulo $2\Delta$, i.e., they are
guaranteed to have intervals that quantize to the same value. If
$2c_p\epsilon\le l-2i\Delta\le \Delta$ for some $i$ we consider the
projection of the balls and the two points, one from each projection,
closest to each other. If these, which have distance $l-2c_p\epsilon$
modulo $2\Delta$, are inconsistent, then the two balls are guaranteed
to be inconsistent. Similarly, if $0 \le l-(2i+1)\Delta\le
\Delta-2c_p\epsilon$ for some $i$ we consider the projection of the
balls and the two points, one from each projection, farthest from each
other. If these, which have distance $l+2c_p\epsilon$ modulo
$2\Delta$, are inconsistent, then the two balls are guaranteed to be
inconsistent.  Since, given $l$, the dither distributes the centers of
the balls uniformly within the quantization intervals, the probability
that we cannot guarantee consistency can be bounded in a manner
similar to~\eqref{eq:projected_cons}.
\begin{align}
  P(\mathrm{no~guarantee}|l) &\le\left\{
    \begin{array}{rl}
      1,&\mbox{if~} |l-2i\Delta|\le 2c_p\epsilon
      \\ \frac{\Delta+2c_p\epsilon-l-2i\Delta}{\Delta},&\mbox{if~}
      2c_p\epsilon \le l-2i\Delta\le
      \Delta\\ \frac{l+2c_pe-(2i+i)\Delta}{\Delta},&\mbox{if~} 0 \le
      l-(2i+1)\Delta\le \Delta-2c_p\epsilon,\\
    \end{array}
    \right.
    \label{eq:no_guarantee}
\end{align}
for some integer $i$. The shape of this upper bound is shown in
Fig.~\ref{fig:ball_quantization}(b). Note that the right hand side
of~\eqref{eq:no_guarantee} can be expressed in terms of $P(q=q'|l)$
from~\eqref{eq:projected_cons} to produce
\begin{align*}
  P\left(\mathrm{no~guarantee}|d\right)&\le
  \min\left\{P\left(q=q'|d\right)+\frac{2c_p\epsilon}{\Delta},1\right\}\\
  &\le P\left(q=q'|d\right)+\frac{2c_p\epsilon}{\Delta}.
\end{align*}
Thus we can upper bound the probability of inconsistent measurements
due to either a large ball projection interval or due to unfavorable
projection of the ball centers using the union bound.
\begin{align*}
  P\left(\exists~\vv\in\ball{\epsilon}{\vx},\vv'\in\ball{\epsilon}{\vx'},
  \mathrm{~s.t.~}q_{\vv}=q_{\vv'}|d\right)&\le P\left(\mbox{no
    guarantee}|d\right)+P\left(\|\phi\|_2\ge c_p\right)\\ &\le
  P\left(q=q'|d\right)+\frac{2c_p\epsilon}{\Delta}+
  \gamma\left(\frac{K}{2},\left(\frac{c_p}{2\sigma}\right)^2\right),
\end{align*}
where $q_{\vv}$ and $q_{\vv'}$ are the quantization values of \vv, and
$\vv'$, respectively. This proves Lemma~\ref{th:single_balls}.

\subsection{Consistency Of $M$ Measurements For All Signals In The Space}
\label{sec:main_result}
To determine the overall quantization performance, we consider bounded
norm signals \vx\ in a $K$ dimensional signal space. Without loss of
generality, we assume $\|\vx\|_2\le 1$, and denote the set of all such
signals using $\sS=\left\{\vx\in\Reals^K, \|\vx\|_2\le 1\right\}$. To
consider all the points in \sS\ we construct a covering using
$\epsilon$-balls, such that any signals in $\sS$ belongs to at least
one such ball. The minimum number of balls required to cover a signal
set is the covering number of the set. For the unit ball in $K$
dimensions, the covering number is $C_{\epsilon}\le (3/\epsilon)^K$
$\epsilon$-balls~\cite{BarDavDeV::2008::A-Simple-Proof}.

Next, we consider all pairs of balls $(\ball{\epsilon}{\vx},
\ball{\epsilon}{\vx'})$, such that $\|\vx-\vx'\|_2\ge d$. The number
of those is upper bounded by the total number of pairs of
$\epsilon$-balls we can form from the covering, independent of the
distance of their centers, namely $\binom{C_{\epsilon}}{2}\le
C_{\epsilon}^2$ pairs. The probability that at least one pair of
vectors, one from each ball has $M$ consistent measurements is upper
bounded by
\begin{align*}
  P(M\mathrm{~measurements~consistent}|d)&=
  P(\exists~\vv\in\ball{\epsilon}{\vx},\vv'\in\ball{\epsilon}{\vx'},
  \mathrm{~s.t.~}\vq_{\vv}=\vq_{\vv'}|d)\\&\le
  P(\exists~\vv\in\ball{\epsilon}{\vx},\vv'\in\ball{\epsilon}{\vx'},
  \mathrm{~s.t.~}q_{\vv}=q_{\vv'}|d)^M
\end{align*}
Thus, the probability that there exists at least one pair of balls
that contains at least one pair of vectors, one from each ball, that
quantize to $M$ consistent measurements can be upper bounded using the
union bound
\begin{align*}
  P(\exists~\vx,\vx'\in\sS,\|\vx-\vx'\|_2>d~\mbox{s.t.}~\vq=\vq')
  &\le\left(\frac{3}{\epsilon}\right)^{2K}P(M\mathrm{~measurements~consistent}|d)
\end{align*}

It follows that the probability that we cannot guarantee inconsistency
for all vectors with distance greater than $d$ is upper bounded by
\begin{align*}
  P(\exists~\vx,\vx'\in\sS,\|\vx-\vx'\|_2>d~\mbox{s.t.}~\vq=\vq')&\le
  \left(\frac{3}{\epsilon}\right)^{2K}\left(
  P(q=q'|d)+\frac{2c_p\epsilon}{\Delta}+
  \gamma\left(\frac{K}{2},\left(\frac{c_p}{2\sigma}\right)^2\right)\right)^M\\ &\le
  \left(\frac{3}{\epsilon}\right)^{2K}\left(
  \frac{1}{2}+\frac{1}{2}e^{-\left(\frac{\pi\sigma
      d}{\sqrt{2}\Delta}\right)^2}+\frac{2c_p\epsilon}{\Delta}+
  \gamma\left(\frac{K}{2},\left(\frac{c_p}{2\sigma}\right)^2\right)\right)^M.
\end{align*}
Picking $\sigma=\frac{1}{\sqrt{K}},~\epsilon=\frac{\Delta r_1}{2c_p}$, and
$\Delta=\frac{dr_2}{\sqrt{K}}$ for some ratios $r_1,r_2>0$ we obtain
\begin{align*}
  P(\exists~\vx,\vx'\in\sS,\|\vx-\vx'\|_2>d~\mbox{s.t.}~\vq=\vq') &\le
  \left(\frac{\sqrt{K}}{d}\frac{6c_p}{r_1r_2}\right)^{2K}\left(
  \frac{1}{2}+\frac{1}{2}e^{\frac{-\pi^2
      }{2r_2^2}}+r_1+
  \gamma\left(\frac{K}{2},\frac{c_p^2K}{4}\right)\right)^M,
\end{align*}
By setting $c_p$ arbitrarily large, and $r_1$ and $r_2$ arbitrarily
small, we can achieve
\begin{align*}
  P(\exists~\vx,\vx'\in\sS,\|\vx-\vx'\|_2>d~\mbox{s.t.}~\vq=\vq') &\le
  \left(\frac{c_o\sqrt{K}}{d}\right)^{2K}\left(
  c_r\right)^M,
\end{align*}
where $c_o=6c_p/r_1r_2$ increases as $c_r$ decreases, and $c_r$ can be
any constant arbitrarily close to 1/2. This proves
Thm.~\ref{th:main}. Corollary~\ref{th:exp_decay} follows trivially.

For example, to make this result concrete, if $K>8$ we can pick
$c_p=2,~\epsilon=\frac{\Delta}{20}$, and $\Delta=\frac{d}{\sqrt{K}}$
to obtain:
\begin{align*}
  P(\exists~\vx,\vx'\in\sS,\|\vx-\vx'\|_2>d~\mbox{s.t.}~\vq=\vq')&\le\left(\frac{60\sqrt{K}}{d}\right)^{2K}\left(\frac{3}{4}\right)^M=
  e^{2K\log\left(\frac{60\sqrt{K}}{d}\right)-M\log\left(\frac{4}{3}\right)}.
\end{align*}

We should remark that the choice of parameters $r_1,r_2$ at the last
step---which also determines the design of the precision parameter
$\Delta$---influences the decay rate of the error, at a trade-off with
the leading constant term. While we can obtain a decay rate
arbitrarily close to $1/2$, we will also force the leading term
$(c_o\sqrt{K}/d)^{2K}$ to become arbitrarily large. As mentioned
before, the decision to decrease $\Delta$ should be done at design
time. Furthermore, decreasing $\Delta$ can be difficult in certain
practical hardware implementations.

The $\sqrt{K}$ factor is consistent with scalar quantization of
orthonormal basis expansions. Specifically, consider the orthonormal
basis expansion of the signal, quantized to $B$ bits per coefficient
for a total of $KB$ bits. The worst-case error per coefficient is
$2^{-(B-1)}$ and, therefore, the total worst-case error is
$2^{-(B-1)}\sqrt{K}$.

To better understand the result, we examine how many bits we require
to achieve the same performance as fine scalar quantization of
orthonormal basis expansions.  To provide the same error guarantee we
set $d=2^{-(B-1)}\sqrt{K}$. Using Corollary~\ref{th:exp_decay}, to
achieve this guarantee with probability $P_0$ we require
\begin{align*}
  2^{-(B-1)}\sqrt{K}&\ge\frac{c_o\sqrt{K}}{P_0^{\frac{1}{2K}}}
  \left(c_r\right)^{\frac{M}{2K}}\\ \Rightarrow \frac{M}{K} &\ge
  \left.2\left(B\log 2
  +\log\frac{c_o}{2P_0^{\frac{1}{2K}}}\right)\right/\log\left(1/c_r\right).
\end{align*}
Thus the number of bits per dimension $M/K$ required grows linearly
with the bits per dimension $B$ required to achieve the same error
guarantee in an orthonormal basis expansion. The oversampled approach
asymptotically requires $2\log(2)/\log(1/c_r)$ times the number of
bits per dimension, compared to fine quantization of orthonormal basis
expansions, an overhead which can be designed to be arbitrarily close
to 2 times. For our example $c_r=3/4$, $2\log(2)/\log(1/c_r)\approx
4.82$. Although this penalty is significant, it is also significantly
improved over classical scalar quantization of oversampled expansions.

\section{Quantization Universality and Signal Models}
\label{sec:side_info}
\subsection{Universality and Side Information}
One of the advantage of our approach is its universality, in the sense
that we did not use any information on the signal model in designing
the quantizer. This is a significant advantage of randomized sampling
methods, such as Johnson-Lindenstrauss embedding and Compressive
Sensing~\cite{johnson1984extensions,DasGup::1999::An-Elementary-Proof,bib:Donoho04A,bib:CandesCS06}.
Additional information about the signal can be exploited in the
reconstruction to improve performance.

The information available about the signal can take the form of a
model on the signal structure, e.g., that the signal is sparse, or
that it lies in a
manifold~\cite{bib:CandesCS06,lu2008theory,blumensath2009sampling,EldMis::2009::Robust-recovery,baraniuk2010model,baraniuk2009random}. Alternatively,
we might have prior knowledge of an existing signal that is very
similar to the acquired one~(e.g., see~\cite{RWBV10}). This
information can be incorporated in the reconstruction to improve the
reconstruction quality. It is expected that such information can allow
us to provide stronger guarantees for the performance of our
quantizer.

We incorporate side information by modifying the set \sS\ of signals
of interest. This set affects our performance through the number of
$\epsilon$-balls required to cover it, known as the covering number of
the set. In the development above, for $K$-dimensional signals with
norm bounded by 1, covering can be achieved by
$C_{\epsilon}=(3/\epsilon)^K$ balls. The results we developed,
however, do not rely on any particular covering number expression. In
general, any set \sS\ can be quantized successfully with probability
\begin{align*}
  P(\exists~\vx,\vx'\in\sS,\|\vx-\vx'\|_2>d~\mbox{s.t.}~\vq=\vq')&\le
  C_{3d/c_o\sqrt{K}}^\sS\left(c_r\right)^M,
\end{align*}
where $C_{\epsilon}^\sS$ denotes the covering number of the set of
interest \sS\ as a function of the ball size $\epsilon$, and $c_o,c_r$
are as defined above.

This observation allows us to quantize known classes of signals, such
as sparse signals or signals in a union of subspaces. All we need for
this characterization is an upper bound for the covering number of the
set (or its logarithm, i.e., the Kolmogorov $\epsilon$-entropy of the
set~\cite{Kolomogorv61}). The underlying assumption is the same as
above: that the reconstruction algorithm selects a signal in the set
\sS\ that is consistent with the quantized measurements. 

The Kolmogorov $\epsilon$-entropy of a set provides a lower bound on
the number of bits necessary to encode the set with worst case
distortion $\epsilon$ using vector quantization. To achieve this rate,
we construct the $\epsilon$-covering of the set and use the available
bits to enumerate the centers of the $\epsilon$-balls comprising the
covering. Each signal is quantized to the closest $\epsilon$-ball
center, the index of which is used to represent the signal. 
While the connection with vector quantization is well understood in
the literature, the results in this paper provide, to our knowledge,
the first example relating the Kolmogorov $\epsilon$-entropy of a set
and the achievable performance under scalar
quantization. Specifically, using a similar derivation to
Cor.~\ref{th:exp_decay}, the number of bits sufficient to guarantee
worst-case distortion $d$ with probability greater than $1-P_0$ is
\begin{align}
  M\ge \frac{\log C_{3d/c_o\sqrt{K}}^\sS+\log \frac{1}{P_0}}{\log
    \frac{1}{c_r}},
\end{align}
where $\log C_{3d/c_o\sqrt{K}}^\sS$ is the $\epsilon$-entropy for
$\epsilon=3d/c_o\sqrt{K}$. Aside from constants, there is a $\sqrt{K}$
penalty over vector quantization in our approach, consistent with the
findings in Sec.~\ref{sec:main_result}.

In the remainder of this section we examine three special cases:
Compressive Sensing, signals in a union of subspaces, and signals with
a known similar signal as side information.

\subsection{Quantization of Sparse Signals}
Compressive Sensing, one of the recent developments in signal
acquisition technology, assumes that the acquired signal $\vx$
contains few non-zero coefficients, i.e., is sparse, when expressed in
some basis. This assumption significantly reduces the number of
measurements required for acquisition and exact
reconstruction~\cite{bib:CandesCS06,bib:Candes04A,bib:Candes04C,bib:Donoho04A}.
However, when combined with scalar quantization it can be shown that
CS measurements are quite inefficient in terms of their
rate-distortion trade-off~\cite{bib:BoufounosDCC07}. The cause is
essentially the same as the cause for the inefficiency of oversampling
in the case of non-sparse signals: sparse signals occupy a small
number of subspaces in the measurement space. Thus, they do not
intersect most of the available quantization points. The proposed
quantization scheme has the potential to significantly improve the
rate-distortion performance of CS.

Compressive Sensing examines $K$-sparse signals in an $N$-dimensional
space. Thus the signal acquired contains up to $K$ non-zero
coefficients and, therefore, lies in a $K$-dimensional subspace out of
the $\binom{N}{K}$ such subspaces. Since each of the subspaces can be
covered with $(3/\epsilon)^K$ balls, and picking
$\sigma=\frac{1}{\sqrt{N}},~\epsilon=\frac{\Delta r_1}{2c_p}$, and
$\Delta=\frac{dr_2}{\sqrt{N}}$, the probabilistic guarantee of reconstruction
becomes
\begin{align*}
  P(\exists~\vx,\vx'\in\sS,\|\vx-\vx'\|_2>d~\mbox{s.t.}~\vq=\vq')&\le
  \binom{N}{K}^2\left(\frac{c_o\sqrt{N}}{d}\right)^{2K}\left(c_r\right)^M\\ &\le
  \left(\frac{eN^{3/2}}{K}
  \frac{c_o}{d}\right)^{2K}\left(c_r\right)^M\\ &\le
  e^{2K\log\left(\frac{eN^{3/2}}{K}\frac{c_o}{d}\right)-M\log(1/c_r)}
\end{align*}
which decays exponentially with $M$, as long as $M=\Omega\left(K\log
N-K\log\left(Kd\right)\right)=\Omega\left(K\log
\left(N/Kd\right)\right)$, similar to most Compressive Sensing
results. The difference here is that there is an explicit
rate-distortion guarantee since $M$ represents both the number of
measurements and the number of bits used.

\subsection{Quantization of Signals in a Union of Subspaces}
A more general model is signals in a finite union of
subspaces\cite{lu2008theory,blumensath2009sampling,EldMis::2009::Robust-recovery,baraniuk2010model}. Under
this model, the signal being acquired belongs to one of $L$
$K$-dimensional subspaces. In this case the reconstruction guarantee
becomes
\begin{align*}
  P(\exists~\vx,\vx'\in\sS,\|\vx-\vx'\|_2>d~\mbox{s.t.}~\vq=\vq')&\le
  L^2\left(\frac{c_o\sqrt{N}}{d}\right)^{2K}\left(c_r\right)^M\\
  & \le e^{2\log L+2K\log\left(\frac{c_o\sqrt{N}}{d}\right)-M\log(1/c_r)},
\end{align*}
which decays exponentially with $M$, as long as $M=\Omega(\log
L+K\log(N/d))$. Compressive Sensing is a special case of signals in a
union of subspaces, where $L=\binom{N}{K}$.

This result is in contrast with the analysis on unquantized
measurement for signals in a union of
subspaces~\cite{lu2008theory,blumensath2009sampling,EldMis::2009::Robust-recovery,baraniuk2010model}. Specifically,
these results demonstrate no dependence on, $N$, the size of the
ambient signal space; $O(\log L+K)$ unquantized measurements are
sufficient to robustly reconstruct signals from a union of
subspaces. On the other hand, using an analysis similar
to~\cite{bib:Thao96,bib:BoufounosDCC07} it is straightforward to show
that increasing the rate by increasing the number of measurements
provides only a linear reduction of the error as a function of the
number of measurements, similar to the behavior described
by~\eqref{eq:oversampling_bound}. Alternatively, we can consider the
Kolmogorov $\epsilon$-entropy, i.e., the minimum number of bits
necessary to represent the signal set at distortion $\epsilon$,
without requiring robustness or imposing linear measurements. This is
exactly equal to $\log_2\left(C_\epsilon^\sS\right)$ and suggests that
$O(\log L+K)$ bits are required. Whether the logarithmic dependence on
$N$ exhibited by our approach is fundamental, due to the requirement
for linear measurements, or whether it can be removed by different
analysis is an interesting question for further research.

\subsection{Quantization of Similar Signals}
Quite often, the side information is a known signal $\vx_s$ that is
very similar to the acquired signal. For example, in video
applications one frame might be very similar to the next; in
multispectral image acquisition and compression the acquired signal in
one spectral band is very similar to the acquired signal in another
spectral band~\cite{RWBV10}. In such cases, knowledge of $\vx_s$ can
significantly reduce the number of quantized measurements required to
acquire the new signal.

As an example, consider the case where it is known that the acquired
signal $\vx$ differs from the side information $\vx_s$ by at most
$D\ge\|\vx_s-\vx\|_2$. Thus the acquired signal exists in the $D$-ball
around $\vx_s$, \ball{D}{\vx_s}. Using the same argument as above, we
can construct a covering of a $D$-ball using $(3D/\epsilon)^K$$
\epsilon$-balls. Thus, the distortion guarantee becomes
\begin{align*}
  P(\exists~\vx,\vx'\in\sS,\|\vx-\vx'\|_2>d~\mbox{s.t.}~\vq=\vq')&\le
  \left(\frac{c_oD\sqrt{K}}{d}\right)^{2K}\left(c_r\right)^M.
\end{align*}
If we fix the probability that we fail to guarantee reconstruction
performance to $P_0$, as with Cor.~\ref{th:exp_decay}, the distorition
guarantee we can provide decreases linearly with $D$.
\begin{align*}
  \|\vx-\vx'\|_2\ge\frac{c_oD\sqrt{K}}{P_0^{\frac{1}{2k}}}
  \left(c_r\right)^{\frac{M}{2K}}\Rightarrow \vq\ne\vq'.
\end{align*}

\section{Discussion and open questions}
\label{sec:discussion}
This paper demonstrates universal scalar quantization with exponential
decay of the quantization error as a function of the oversampling rate
(and, consequently, of the bit rate). This allows rate-efficient
quantization for oversampled signals without any need for methods
requiring feedback or joint quantization of coefficients, such as
Sigma-Delta or vector quantization. The framework we develop is
universal and can incorporate side information on the signal, when
available. Our development establishes a direct connection between the
Kolmogorov $\epsilon$-entropy of the measured signals and the
achievable rate vs. distortion performance under scalar quantization.

The fundamental realization to enable this performance is that
continuous quantization regions (i.e., monotonic scalar quantization
functions) cause the inherent limitation of scalar quantizers. Using
non-continuous quantization regions we make more effective use of the
quantization bits. While in this paper we only analyze binary
quantization, it is straightforward to analyze multibit quantizers,
shown in Fig.~\ref{fig:q_fun}(c). The only difference is the
probability $P(q=q'|l)$ that two arbitrary signals produce a
consistent measurement in~\eqref{eq:projected_cons} and
Fig.~\ref{fig:consistent}(b). The modified function should be equal to
zero in the intervals
$[(2^Bi+1)\Delta,(2^B(i+1)-1)\Delta],i=0,1,\ldots$, and equal
to~\eqref{eq:projected_cons} everywhere else. The remaining derivation
is identical to the one we presented. We can conjecture that careful
analysis of the multibit case should present an exponential decay
constant $c_r\gtrsim 1/2^B$, which can reach that lower bound
arbitrarily close.

One of the issues not addressed in this work is practical
reconstruction algorithms. Reconstruction from the proposed sampling
scheme is indeed not straightforward. However, we believe that our
work opens the road to a variety of scalar quantization approaches
which can exhibit practical and efficient reconstruction
algorithms. One approach is to use the results in this paper
hierarchically, with a different scaling parameter $\Delta$ at each
hierarchy level, and, therefore, different reconstruction accuracy
guarantees. The parameters can be designed such that the
reconstruction problem at each level is a convex problem, therefore
tractable. This approach is explored in more detail
in~\cite{bib:BoufSAMPTA2011}. We defer discussion of other practical
reconstruction approaches to future work.

A difficulty in implementing the proposed approach is that the
precision parameter $\Delta$ is tightly related to the hardware
implementation of the quantizer. It is also critical to the
performance. If the hardware is not precise enough to scale $\Delta$
and produce a fine enough quantization function $Q(x)$, then the
asymptotic performance of the quantizer degrades. This is generally
not an issue in software implementations, e.g., in compression
applications, assuming we do not reach the limits of machine
precision.

The precision parameter $\Delta$ also has to be designed in advance to
accommodate the target accuracy. This might be undesirable if the
required accuracy of the acquisition system is not known in advance,
and we hope to decide the number of measurements during the system's
operation, maybe after a certain number of measurements has already
been acquired with a lower precision setting. One approach to address
this issue is to hierarchically scale the precision parameter, such
that the measurements are more and more refined as more are
acquired. The hierarchical quantization discussed
in~\cite{bib:BoufSAMPTA2011} implements this approach.

Another topic worthy of further research is performance in the
presence of noise. Noise can create several problems, such as
incorrect quantization bits. Even with infinite quantization
precision, noise in an inescapable fact of signal acquisition and
degrades performance. There are several ways to account for noise in
this work. One possibility is to limit the size of the precision
parameter $\Delta$ such that the probability the noise causes the
measurement to move by more than $\Delta$ can be safely ignored. This
will limit the number of bit flips due to noise, and should provide
some performance guarantee. It will also limit the asymptotic
performance of the quantizer. Another possibility is to explore the
robust embedding properties of the acquisition process, similar
to~\cite{bib:JLBB_BeSE}. More precise examination is an open question,
also for future work.

An interesting question is the ``democratic'' property of this
quantizer, i.e. how well the information is distributed to each
quantization
bit~\cite{CalDau::2002::The-pros-and-cons,davenport-simple,bib:LBDB_ACHA11}.
This is a desirable property since it provides robustness to erasures,
something that overcomplete representations are known
for~\cite{bib:Goyal01,boufounos2008causal}. Superficially it seems
that the quantizer is indeed democratic. In a probabilistic sense, all
the measurements contain the same amount of information. Similarities
with democratic properties in Compressive
Sensing~\cite{davenport-simple} hint that the democratic property of
our method should be true in an adversarial sense as well. However, we
have not attempted a proof in this paper.

Last, we should note that this quantization approach has very tight
connections with locality-sensitive hashing (LSH) and $\ell_2$
embeddings under the hamming distance~(e.g., see \cite{Andoni08LSH}
and references within). Specifically, our quantization approach
effectively constructs such an embedding, some of the properties of
which are examined in~\cite{raginsky2009locality}, although not in the
same language.  A significant difference is on the objective. Our goal
is to enable reconstruction, whereas the goal of LSH and randomized
embeddings is to approximately preserve distances with very high
probability. A rigorous treatment of the connections of quantization
and LSH is quite interesting and deserves a publication of its own. A
preliminary attempt to view LSH as a quantization problem is performed
in~\cite{brandt2010transform}.

\bibliographystyle{IEEEtran} 
\bibliography{IEEEabrv,biblio}
\end{document}